\documentclass[a4paper,11pt]{article}
\usepackage{jcappub} 

\usepackage{calrsfs}
\usepackage{multirow}
\DeclareMathAlphabet{\pazocal}{OMS}{zplm}{m}{n}
\newcommand{\Ll}{\mathcal{L}}
\newcommand{\Lp}{\pazocal{L}}
\usepackage{lineno}
\usepackage[normalem]{ulem}

\title{\boldmath BAO cosmology in non-spatially flat  background geometry from BOSS+eBOSS and lessons for future surveys}

\author[a]{Santiago Sanz-Wuhl}
\author[b,c]{H\'ector Gil-Mar\'in\footnote[1]{Corresponding author}}
\author[a]{Antonio J. Cuesta}
\author[b,d]{\, Licia Verde}
\affiliation[a]{Departamento de F\'isica, Universidad de C\'ordoba,\\
Campus Universitario de Rabanales, Ctra. N-IV Km.~396,
E-14071 C\'ordoba, Spain}
\affiliation[b]{Institut de Ci\`encies del Cosmos (ICCUB), Universitat de Barcelona (UB), C.~Mart\'i i Franqu\'es, 1, 08028 Barcelona, Spain.}
\affiliation[c]{Departament de F\'isica Quàntica i Astrof\'isica (FQA), Universitat de Barcelona (UB),  C.~Mart\'i i Franqu\'es, 1, 08028 Barcelona, Spain}
\affiliation[d]{Instituci\'o Catalana de Recerca i Estudis Avan\c cats, ICREA,\\
Passeig Llu\'is Companys 23, E-08010 Barcelona, Spain}

\emailAdd{f82sawus@uco.es, hectorgil@icc.ub.edu, ajcuesta@uco.es, liciaverde@icc.ub.edu}

\abstract{We study the impact of the assumption of a non-flat fiducial cosmology on the measurement,  analysis and interpretation of BAO distance variables, along and across the line-of-sight. The assumption about cosmology enters in the choice of the base template, as well as on the transformation of tracer's redshifts into distances (the catalog cosmology): here we focus on the curvature assumption, separately and jointly, on both.

We employ BOSS and eBOSS publicly available data and show that for the statistical precision of this data set, distance measures and thus cosmological inference are robust to assumptions about curvature both of the template and the catalog. Thus the usual assumptions of flat fiducial cosmologies (but also assumptions of non-flat cosmologies) do not produce any detectable systematic effects.
For forthcoming large-volume surveys, however, small but appreciable residual systematic shifts can be generated which may require some care. These are mostly driven by the choice of catalog cosmology if it is significantly different from true cosmology.
In particular, the catalog (and template) cosmology should be chosen, possibly iteratively, in such a way that the recovered BAO scaling variables are sufficiently close to unity. At this level of precision, however, other previously overlooked effects become relevant, such as a mismatch between the sound horizon as seen in the BAO and the actual sound horizon in the early Universe. If unaccounted for, such effect may be misinterpreted as cosmological and thus bias the curvature (and cosmology) constraints. We present best practices to avoid this.
}

\begin{document}
\maketitle
\flushbottom

\section{Introduction}
\label{sec:intro}
The standard model of cosmology, the $\Lambda$CDM model, assumes a  spatially flat Universe, cosmological constant-dominated, with cold dark matter and baryons as the components that cluster under gravity\footnote{While neutrinos do cluster under gravity,  non-relativistic neutrinos contribute to clustering in a sub-dominant way so much so that in its simplest incarnation the $\Lambda$CDM model assumes massless neutrinos.}. The spectacular success of this model over the past three decades was driven by an avalanche of data, among them, the Cosmic Microwave Background (CMB) anisotropies missions,  Wilkinson Microwave Anisotropy Probe (WMAP, \citep{Hinshaw2012}) and Planck \citep{Planck18}, the spectroscopic galaxy surveys such as Sloan Digital Sky Survey Baryon Oscillation Spectroscopic Survey (SDSS BOSS, \citep{alametal17}) and extended BOSS (SDSS eBOSS, \citep{alametal21}), as well as photometric galaxy surveys such as the Dark Energy Survey (DES,\cite{DES}) and the KiloDegree galaxy Survey (KiDS,\cite{kids1,kids2}). The Baryon Acoustic Oscillation (BAO) peak, has been one of the main probes that has helped to consolidate this standard model. Ongoing surveys, such as the Dark Energy Spectroscopic Instrument (DESI, \cite{desi,moonetal:2023}) and the EUCLID mission \cite{euclid}, will exploit BAO to push the model to an unprecedented level of accuracy. 

The parameters of the model are now constrained with exquisite precision, in some cases at
better than 1\% level.
Gradually, the focus of efforts in the interpretation of new data has shifted from further constraining the model parameters, to precision tests of the model itself and search for new physics beyond it. This is usually done by extending the model with additional parameters which describe well-defined deviations.
One such additional parameter which has received renewed attention recently is the background spatial curvature, $\Omega_k$ \citep{Valentino2019,Glanvilleetal,Dhawanetal21,Liuetal22,Vagnozzietal20a,Vagnozzietal20b,Yangetal22,Handley2019}. 
A convincing detection of a non-zero spatial curvature would have profound implications for inflationary theories and challenge entire families of inflationary models (see \citep{Linde2007} and refs. therein, see also \cite{guthetal12,guthetal14}). 

The constraining power of CMB primary anisotropy observations is affected by the well-known geometrical degeneracy \cite{Bond1997,Efstathiou1998,kamionkowski94}:
without imposing spatial flatness, closed geometries provide good fits to the data. Positive curvature models have been claimed to be preferred over the standard model (in the Bayesian sense)  by the latest CMB data, although it is not completely clear what is the role of the adopted prior on the parameters being marginalized in the posterior calculation and the role of the prior on the curvature parameter itself \cite{Vagnozzietal20b,plancklike,Efstathiou_Gratton21}.

Secondary effects such as lensing can ease the geometric degeneracy.
The geometric degeneracy is broken by combining CMB data with late-time universe geometric measurements and in particular with galaxy, quasar and Ly-$\alpha$ BAO data (see for eg., \cite{cuesta_baodr12,gilmarin_baodr12}). Observations of the standard ruler, provided by the imprint of the sound horizon at radiation drag on the clustering of large-scale structure tracers, help constrain the late-time expansion history of the Universe, $H(z)$. Since the shape of $H(z)$ depends on how the different contents of the Universe evolve with redshift, constraining $H(z)$ also helps to constrain the space-time geometry.

Any galaxy, quasar and  Ly-$\alpha$ maps are obtained as a function of two angular components which determine the position of an object in the sky (typically, the right ascension and declination angles), and the redshift. To compute the summary statistics from the map (i.e., the $n$-point correlators) we must first transform the redshift into a radial distance. This transformation must be done by assuming a reference cosmology which is often referred to in the literature as `fiducial cosmology' \cite{alametal17,alametal21}. In this paper, we refer to this assumed cosmology as `catalog cosmology' (as it is the fiducial cosmology to construct the catalog in comoving coordinates). On the other hand, most of the state-of-the-art BAO analyses are template-based: a set template for the  BAO oscillations is assumed and the BAO features on the data are quantified relative to that reference template. The chosen cosmology for that template is also referred to in the literature as  `fiducial cosmology'.  In this paper, we refer to this assumed cosmology as `template cosmology' (as it is the fiducial cosmology to construct the model for the summary statistics starting from a given template). In almost all BAO analyses, both template and catalog cosmology are chosen to be the same for practical reasons, but this is just a matter of choice.

Usually, such template and catalog cosmologies are consistent with the standard (flat) $\Lambda$CDM model,  with curvature parameter  $\Omega_k=0$, favoured by observations of the CMB anisotropies. Such a choice has raised some questions about the validity of the BAO analyses: whether the preference for flatness provided by  BAO constraints when combined with CMB is a circular argument, which stems from the initial assumptions of both catalog or template assumed cosmologies being flat. 

While BAO analyses should in principle be insensitive to the particular choice of template or catalog cosmologies, this should be demonstrated to be the case at the precision level reached by state-of-the-art BAO data.
The main goal of this paper is to perform this check for deviations from spatial flatness.

Previous papers have already investigated such an effect, but not for $\Omega_k\neq0$ fiducial choices on the clustering of galaxies. In \cite{Carteretal2020} the authors study the impact of the choice of the template and catalog cosmologies when fitting a suite of flat $w$CDM cosmologies, finding small trends on the BAO parameters with the true parameters of the simulations employed. In \cite{helion2020} the authors explore the impact of non-zero curvature fiducial choices on the BAO parameters for the eBOSS Ly-$\alpha$ analysis, testing variations of the Hubble parameter $h$ and $\Omega_k$ while keeping fixed the physical densities of baryons and cold dark matter,  $\omega_b$ and $\omega_c$, to the fiducial choices. Their results support the idea that, at least for the eBOSS precision, the assumption of flatness has no measurable impact. 

Inspired by these works, we explore the impact of the catalog and template cosmology choice in the clustering of galaxies on the inference of cosmological parameters for  BOSS+eBOSS, and for a DESI-like survey 10 times larger the volume of the full BOSS+eBOSS.
This paper is structured in two parts. In the first part we answer the question: ``Are current spatial curvature bounds from BAO data biased by the assumption of a flat fiducial cosmology?". In the second part of the paper we address instead the following: ``For the increased statistical precision  of forthcoming surveys, what are the residual systematics induced by the (arbitrary) choice of curvature parameter in  the fiducial model? and  how should these be accounted for and mitigated in the analysis?"

In \S~\ref{sec:bg} we briefly introduce the BAO methodology based on the fixed template approach and we present the data and simulations we employ in this paper. In \S~\ref{sec:results1} we
quantify the sensitivity of the flatness constraints on the choice of fiducial model for the BAO eBOSS data and for the full BOSS and eBOSS samples combined.
In \S~\ref{sec:results2} we explore the potential residual systematics of the main BAO pipeline employing a data vector which mimics the precision of a survey 10 times the combined BOSS+eBOSS volume. Finally in \S~\ref{sec:conclusions} we summarize the main conclusions of this work and their implications for present and future surveys. 

\section{Background and methodology}\label{sec:bg}

We start by briefly reviewing the basic methodology used for extracting the BAO information from a spectroscopic galaxy map in \S \ref{sec:review} and to perform cosmological inference in \S \ref{sec:cosmoinference}. The methodology used matches closely the one used for BOSS and eBOSS analyses (see e.g., \cite{gilmarin20}).  The data and simulations used are presented in \S \ref{sec:data}.

The specific methodology employed in this paper with the specific tests performed are presented in \S \ref{sec:methodology}.

 In what follows the symbol ${\bf \Omega}_i$ refers to a suite of values for the cosmological parameters of a given model. Given ${\bf \Omega}_i$,  quantities like Hubble distance, angular diameter distance (see Appendix \ref{sec:appendixA} for expressions), linear matter power spectrum, sound horizon length at radiation drag, etc., can readily be computed.  
 
\subsection{Review of template-based BAO analysis}
\label{sec:review}
The main role of the BAO feature in cosmology is that it can be used as a standard ruler. Under the cosmological principle,  statistical isotropy is assumed. However, an incorrect choice of cosmology in converting redshift into distances breaks isotropy. In fact, it distorts differently the comoving BAO feature scale across the line-of-sight and along the line-of-sight (as the integral across redshift is weighted wrongly). Hence the correct cosmology can be identified as the one keeping longitudinal and transversal BAO scales equal (as per the cosmological principle). This is the well-known Alcock-Paczynski effect \cite{AP}. In addition, when several redshift bins are employed in the analysis, the consistency of the standard ruler across redshifts \footnote{A very small shift is expected due to non-linear effects, which also depend on redshift. However, this shift is typically tiny compared to the statistical precision of the current state-of-the-art BAO measurements \cite{blasetal}.} can be used to remove cosmological parameter degeneracies.

Although this image helps to visualize and understand how the cosmological information is enclosed in a galaxy map, it is not too useful to understand how the current state-of-the-art BAO pipelines work, essentially because of two practical reasons. First, the transformation from the map (in terms of angles and redshift) to the summary statistics (the correlation function, $\xi(r)$, or the power spectrum, $P(k)$ as it will be used in this work) is computationally too expensive to be included within a typical inference pipeline. If included, such transformation would have to be iteratively repeated  $\sim10^5$ times. Second, when only large-scale structure data is used, it is not possible to calibrate absolute distances, and the BAO scale can only be expressed in units of the physical size of the horizon scale, $r_d$. 

The first issue can be tackled by choosing a known but arbitrary reference cosmology to convert redshift into distances, the `catalog cosmology'. Doing so introduces an anisotropic signal and therefore the apparent scales of the  BAO signal along and across the line-of-sight are expected to be different.
By quantifying this anisotropic signal it is possible to estimate how different the fiducial cosmology choice is from the actual one, making cosmological inference possible.
The second issue can be addressed either by calibrating the sound horizon at radiation drag $r_d$ (relying on assumptions about the early time physics governing the pre-recombination Universe) or by simply not calibrating the length of the ruler (sometimes referred to as uncalibrated BAO, for e.g., \cite{Brieden2h}).

\subsubsection{Scaling parameters}
The two variables, 
\begin{equation}
    q_\parallel(z) = \frac{D_H(z)}{D_H^{\rm c}(z)};\quad\quad\quad q_\perp(z) = \frac{D_M(z)}{D_M^{\rm c}(z)};
\end{equation}
quantify the distortion of the BAO scale along and across the line of sight, respectively. These two variables are connected to the radial Hubble distance, $D_H(z)=c/H(z)$, and the angular diameter distance, $D_M(z)$ (see Appendix \ref{sec:appendixA}). 
Hereafter, the superscript c stands for the `catalog cosmology' value, whereas the quantity without superscript refers to the actual (unknown in case of observations) underlying `true' cosmology. 

The dependence on the length of the standard ruler, $r_d$, and its effect on cosmological inference is usually addressed as follows (this is known as the `fixed template' approach)\footnote{Other approaches not based on the fixed template, e.g., \cite{shifan}, do not have an associated fiducial $r_d$, as the sound horizon is iteratively modified with the inferred `trial' cosmology. However, they still rely on the choice of the catalog cosmology.}.
A template for the linear matter power spectrum is assumed, computed in a given -fixed-reference cosmology: the `template cosmology'. Such a template includes a broadband component (which does not carry any BAO information) and an oscillatory, BAO, component.  

During inference, the sound horizon scale parameter is treated as a free parameter of the model and, in the template, the  BAO peak position is shifted away from that of the template cosmology accordingly. This represents an isotropic shift of the BAO scale: the sound horizon scale is assumed to be the same along and across the line-of-sight, which can be seen just as a change in the choice of units of the BAO size.
 This is implemented as a re-scaling of $q_{\parallel,\perp}$ by the BAO sound horizon scale in the template cosmology, $r_d^{\rm t}$. In this work, the superscript t stands for the `template cosmology'. Taking into account that the angular and radial BAO distances, $D_H$ and $D_M$ are in turn also expressed in terms of the sound horizon scale units of the actual cosmology, $r_d$,  the usual BAO scaling parameters, $\alpha_{\parallel,\perp}$ are defined as\footnote{Note that since the comoving distances in our 3D catalogs are expressed in Mpc$h^{-1}$ $D_H^c$ and $D_M^c$ are commonly given in units of ${\rm Mpc}\,h^{-1}$, with $h$ the underlying true variable, $\alpha_{\parallel,\,\perp}$ do not depend on $h^c$. On the other hand, since the power spectrum template is given in units of ${\rm Mpc}\,{h^t}^{-1}$, it requires that $r_d^t$ to be also given in units of ${\rm Mpc}\,{h^t}^{-1}$. Thus, $\alpha_{\parallel,\perp}$ depend on the normalized expansion history of the catalog cosmology, $E({\bf \Omega}^c)$, the sound horizon scale at the template cosmology, $r_d^t$, expressed in units of the template cosmology, ${\rm Mpc}\,{h^t}^{-1}$. Since $r_d^t$ is commonly given instead in units of Mpc, the $\alpha_{\parallel,\,\perp}$ will require in this case an extra $h/h^t$ factor.},
\begin{equation}
        \alpha_\parallel(z) = \frac{D_H(z)/r_d}{D_H^{\rm c}(z)/r_d^{\rm t}};\quad\quad\quad \alpha_\perp(z) = \frac{D_M(z)/r_d}{D_M^{\rm c}(z)/r_d^{\rm t}}.
        \label{eq:scaling}
\end{equation}

\subsubsection{The template and its multipoles}\label{sec:template}
In practice, the BAO linear template model for the power spectrum  is  written as
 \cite{gilmarin20,beutler17},
\begin{equation}
    P_{\rm BAO}(k,\mu)=B(1+\beta_{\rm eff}\mu^2)P^{\rm sm}_{\rm lin}(k)\left\{1+[\pazocal{O}_{\rm lin}(k)-1]e^{-\frac{1}{2}k^2[\mu^2\Sigma^2_\parallel+(1-\mu^2)\Sigma_\perp^2]}  \right\}
    \label{eq:Pbao}
\end{equation}
where the $B$ and $\beta_{\rm eff}$ are non-BAO nuisance parameters\footnote{These parameters are inspired by the linear bias, $b_1^2=B$ and the parameter from redshift space distortions, $\beta=f/b_1$. However, here they are not given any cosmological interpretation as we are only interested in the BAO signal. 
An analysis where we study the impact of the reference cosmology on these parameters is beyond the scope of this paper.}, $P^{\rm sm}_{\rm lin}$ stands for the (broadband) BAO-smoothed linear power spectrum, $\pazocal{O}_{\rm lin}$ is the BAO component given by the ratio between the linear power spectrum and the BAO-smoothed versions, and $\Sigma_{\parallel}$ and $\Sigma_\perp$ are parameters that regulate the smoothing of the BAO oscillations. They account for the effect of peculiar velocity bulk-flows on the BAO amplitude, along and across the line-of-sight directions, respectively. We have defined $\mu\equiv\cos\theta$, where $\theta$ is the angle with respect to the line of sight direction.   

The sensitivity to anisotropic distortions is usually captured by computing the \textit{monopole}, $P^{(0)}(k)$, and \textit{quadrupole}, $P^{(2)}(k)$, of the galaxy power spectrum with respect to the line-of-sight (although fitting the full angular dependence $P(k,\mu)$ is possible in some cases). 

The multipoles are computed from the template as,
\begin{equation}
    P^{(\ell)}_{\rm BAO}(k) = \frac{2\ell+1}{2}\int_{-1}^{+1} d\mu\, P_{\rm BAO}(k',\mu') {\Lp}_\ell(\mu) + \sum^n_{i=1} A_i^{(\ell)} k^{2-i},
    \label{eq:Pbao2}
\end{equation}
where ${\Lp}_{\ell}$ denotes the Legendre polynomial of degree $\ell$, and $k'$ and $\mu'$ stands for the observed scale and line-of-sight modified by the scaling parameters of Eq.~\ref{eq:scaling},
\begin{equation}
    k'=\frac{k}{\alpha_\perp}\left[1+\mu^2\left(\frac{\alpha_\perp^2}{\alpha_\parallel^2} -1\right) \right]^{1/2} \quad \mu'=\mu\frac{\alpha_\perp}{\alpha_\parallel}\left[1+\mu^2 \left(\frac{\alpha_\perp^2}{\alpha_\parallel^2} -1\right) \right]^{-1/2}.
\end{equation}
Note that additional nuisance parameters $A_i^{(\ell)}$ in Eq~\ref{eq:Pbao2} are added to the multipoles to absorb non-linear effects.
The nuisance parameters $B, A_i^{(\ell)},\beta_{\rm eff}$ (see for eg., section 3.1 of \cite{gilmarin20} for a detailed description) are assumed to not correlate with the scaling parameters $\alpha_{\parallel,\perp}$ in any significant way, or, equivalently, the signal in the broad-band term of the power spectrum can be decoupled completely from the oscillatory BAO part. 

In this work, we choose $n=5$, following the same criteria as previous BOSS works \cite{ross17}. We fix $\Sigma_\parallel=9.4\, {\rm Mpc}h^{-1}$ and $\Sigma_\perp=4.8\, {\rm Mpc}h^{-1}$ following the fiducial choices of eBOSS LRG sample analysis \cite{gilmarin20}. Although there is a weak correlation between $\Sigma_{\parallel,\perp}$ and $\alpha_{\parallel,\perp}$, for the statistical precision of BOSS and eBOSS this effect has been reported to be sub-dominant \cite{gilmarin20,bautista20}.
The 11 nuisance parameters $\{A_i^{(0,2)}, B\}$, are varied independently in the North and South galactic cap portions of the survey (but $\beta_{\rm eff}$ is assumed to be the same across the two survey portions), yielding a total of 23 nuisance parameters per redshift bin.

Finally, when comparing the power spectrum data measured in a realistic survey geometry, the effects of the survey window function need to be taken into account. We follow the standard procedure and convolve the theoretical power spectrum of Eq.~\ref{eq:Pbao2} with the window selection function, $W_\ell^2(s)$ \cite{Wilson} (see also appendix D of \cite{gilmarin20}). These functions, the $W_\ell^2(s)$,  are computed through a pair count of the random catalogs (catalogs with the same radial and angular geometric selection function but with no intrinsic clustering), and therefore are computationally expensive. In this paper, we assume that the effect of the choice of fiducial (mask) cosmology to infer the $W_\ell^2(s)$  from the random catalogs has a negligible impact on the results (for the precision being considered), and therefore we employ in all the cases the same $W_\ell^2(s)$ function, evaluated at the flat fiducial $\Lambda$CDM model. In Appendix~\ref{app:mask} we validate this assumption.

\subsubsection{Inference}
In this approach, cosmological inference is performed via a likelihood analysis of the power spectrum monopole and quadrupole using the modeling presented above: the parameters of cosmological interest are $\alpha_{\parallel}(z_j)$, $\alpha_\perp(z_j)$ where $j$ runs over the catalog redshift bins (or alternatively $D_H(z_j)/r_d$, $D_M(z_j)/r_d$), and the nuisance parameters.

The freedom provided by the fixed template cosmology expression of Eqs.~\ref{eq:Pbao} and \ref{eq:Pbao2} with the scaling parameters $\{\alpha_\parallel,\,\alpha_\perp\}$ and the  11 nuisance parameters $\{A^{(\ell)}_i, B\}$ per survey portion, along with the global parameter $\beta_{\rm eff}$, is sufficient to provide a  good fit to the data given the current statistical power of the state-of-the-art surveys.

However, this approach could in principle introduce subtle biases in the recovered distances and thus cosmological parameters if the catalog and/or template cosmologies are very different --given the statistical errors of the data--  from the actual cosmology of the universe (see e.g., \cite{Carteretal2020} for quantification of this residual effect on flat $w$CDM models).
In section \ref{sec:results2}, we thoroughly examine for the first time this effect for changes produced by varying the curvature parameter  $\Omega_k$ in the assumed template and catalog cosmologies for the galaxy clustering (see \cite{helion2020} for the effect of non-flat $\Lambda$CDM fiducial choices in Lyman-$\alpha$ BAO analyses).

The likelihood analysis requires a 
covariance that encodes the uncertainties in the summary statistics measurements, $P(k)$. The shape and amplitude of the measured $P(k)$ depend on the chosen value for the catalog cosmology, $P(k|{\bf \Omega^{\rm c}})$, therefore the covariance of $P$ must also account for this dependence. Since for BOSS and eBOSS catalogs the covariance is numerically estimated by using fast mocks \cite{ezmocks,patchy}, as long as the power spectrum of these mocks is computed by assuming the same catalog cosmology as the data, the covariance will include this effect. Alternative approaches which may obtain the covariance analytically \cite{wadekar}, should also incorporate this effect when studying the impact of the assumptions of the reference cosmologies in the final inferred parameters. In Appendix~\ref{sec:appB} we show the effect of ignoring the dependence on the fiducial catalog cosmology in the covariance, and how to account for it analytically. 

\begin{figure}[htb]
    \centering
        \includegraphics[width=1\textwidth]{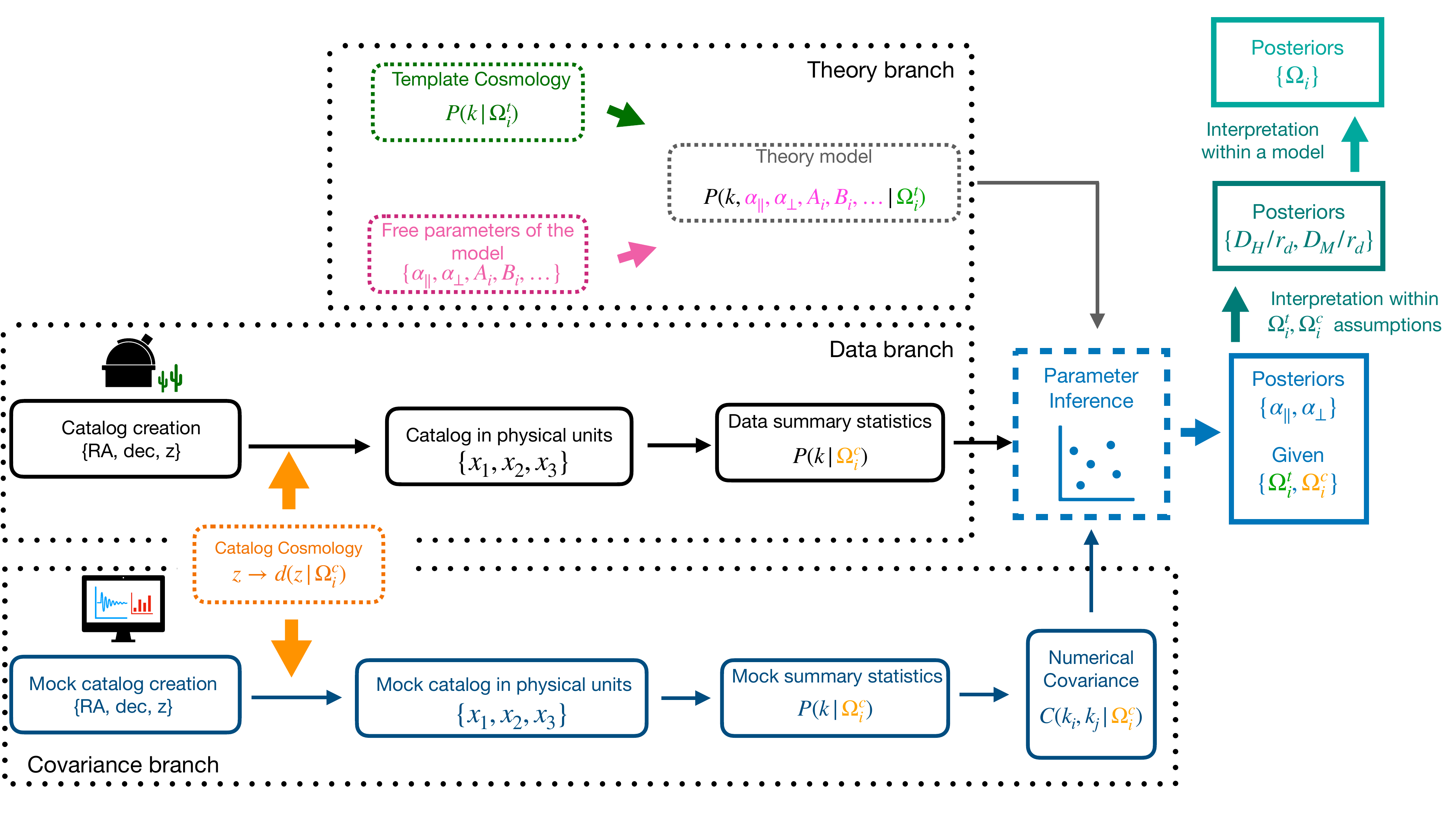}
    \caption{ Schematic representation of the steps and dependencies in the BAO fixed-template pipeline. Inputs for parameter inference come from three branches.  The observations (the catalog creation) are given in angles (right ascension, declination) and redshifts, $z$. A reference catalog cosmology (in orange) needs to be used to convert angles and redshifts into distances (here schematically indicated by Cartesian coordinates $(x_1,x_2,x_3$)) from which a  suitable summary statistics, here, the power spectrum, can be computed (data branches). The same structure needs to be applied to the mock catalogs for generating a numerical covariance matrix (covariance branch). In template-based analyses, a reference template cosmology (in green) is also assumed to compute the template and generate a theory model along with several free parameters (in magenta), as represented in the theory branch. All these three branches converge in the inference parameter box, where the cosmology-related quantities posteriors, $\alpha_{\parallel,\,\perp}$, are obtained. These are later interpreted in terms of the BAO distances given in units of the sound horizon scale, $D_H/r_d$, $D_M/r_d$ which can later be interpreted as parameters of the model, and which are independent of the reference cosmology choices.} 
    \label{fig:sketch}
\end{figure}

Figure~\ref{fig:sketch} schematically summarizes the steps and dependencies in the BAO fixed-template pipeline. Orange symbols (subscript $\mathrm{c}$) and boxes refer to the catalog cosmology, and green (subscript $\mathrm{t}$) to the template cosmology. The free parameters of the model, on which parameter fitting/inference is performed, are in magenta. Parameter inference requires inputs from three `branches': the data branch, the theory branch and the covariance branch (dotted boxes in the figure). The covariance branch illustrates the classic steps for obtaining a numerical covariance from a suite of mock catalogs.

After parameter inference, the $\alpha_\parallel(z)$ and $\alpha_\perp(z)$ posteriors depend on the arbitrary choices of $\bf {\Omega}^{\rm c,t}$ (see eq.~\ref{eq:scaling}), where here ${\bf \Omega}$ represents the full vector of parameters of the model, $\{\Omega_m,\,\Omega_k,\,r_d,\,h\}$.  The $\alpha$ parameters are then interpreted in terms of the BAO distance parameters $D_H(z)$ and $D_M(z)$ in units of $r_d$.
Finally, these distances can be used to provide constraints on the parameters of the cosmological model of choice (i.e., to do inference for the values of parameters $\Omega_i$ of a model).

We will show below that the dependence on the arbitrary choices of ${\Omega}^{\rm c,t}$ has no measurable impact on the inferred cosmology-related quantities for BOSS and eBOSS precision, for reasonable choices of  ${\Omega}^{\rm c,t}$. For future surveys, however, this needs some care (see \S~\ref{sec:results2}).

\begin{figure}[htb]
    \centering
    \includegraphics[width=0.75\textwidth]{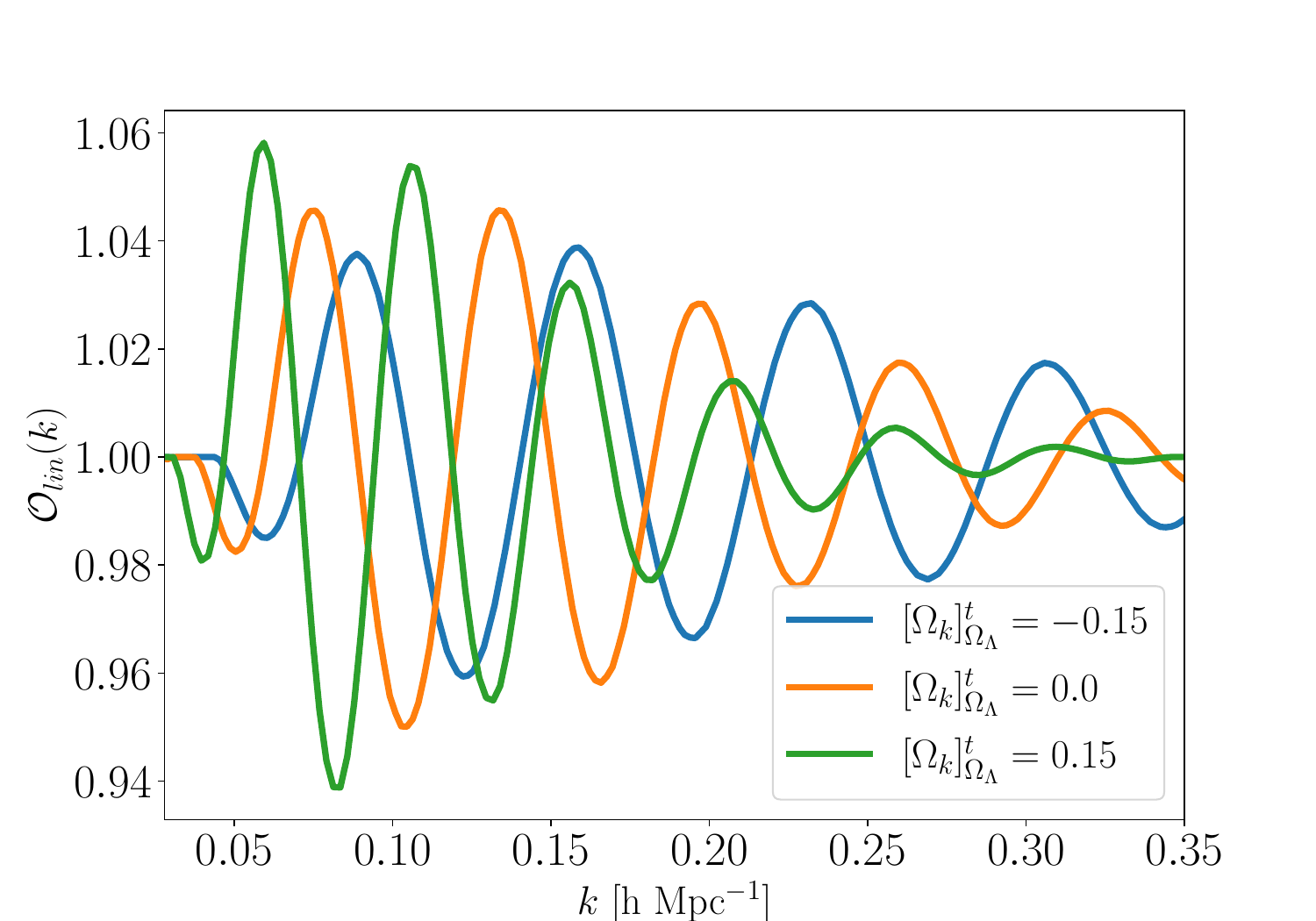}
    \caption{Oscillatory features represented in the $\pazocal{O}_{\rm lin}(k)$ function within Eq.~\ref{eq:Pbao}. The different colors correspond to different values of $\Omega_k$ in the template cosmology obtained as a variation of $\Omega_m$ while keeping $\Omega_\Lambda=0.69$ fixed.  The cold dark matter-to-baryon density ratio is kept fixed at $\Omega_c/\Omega_b=5.43269$,  to limit extreme variations in the   BAO oscillation amplitude. These $\pazocal{O}_{\rm lin}$ functions with non-zero values of $[\Omega_k]^t_{\Omega_\Lambda}$ are used to generate the $P^{(\ell)}_{\rm BAO}(k)$ templates  and to obtain the results of \S~\ref{sec:results1}-\ref{sec:results2} and right panels of Figs.~\ref{fig:ffct} and \ref{fig:cfct}.}
    \label{fig:olin}
\end{figure}

\subsection{Data and simulations}
\label{sec:data}
We employ the publicly available catalogs released by the SDSS BOSS and eBOSS collaborations \cite{web1,web2}. These consist of 1,\,002,\,101 luminous red galaxies (LRG) observed by the SDSS-III BOSS program in the range $0.4<z<0.6$ \cite{Reid17}, 377,\,458 LRG observed by SDSS-III BOSS in combination with SDSS-IV eBOSS program in the range $0.6<z<1.0$ \cite{Ross2020}, and 343,\,708 quasars observed by SDSS-IV program in the range $0.8<z<2.2$ \cite{QSOcat}. These objects are divided into  North and South galactic caps. As it is customary, we consider these to be statistically independent. The total effective volume of these samples is $V_{\rm eff}=2.82\,[{\rm Gpc}\,h^{-1}]^3$.

The power spectrum multipoles are extracted by running \textsc{Rustico} \cite{rustico_github} with different assumptions on the flatness of the background catalog cosmology. To estimate the covariance of the power spectrum multipoles we employ the publicly released mock galaxy catalogs: 2048 realizations of the \textsc{MD-Patchy} mocks \cite{patchy} for BOSS LRGs, and 1000 realizations of the \textsc{EZmocks} for both eBOSS LRGs and quasars \cite{ezmocks}.

 We do not employ reconstruction (see e.g., \cite{alametal21,recon1,recon2} and refs therein). While reconstruction sharpens the BAO signature and mitigates effects of non-linearities thus allowing for more accurate cosmological inference, it introduces an extra layer of complexity and dependence on fiducial cosmology assumptions. The statistics of the ``post-recon" data depend on the fiducial cosmology assumed and quantifying this dependence goes well beyond the scope of this paper. For simplicity and transparency here  we use exclusively non reconstructed (also called  ``pre-recon") data. Moreover most of the signal come from relatively high redshifts where the gains offered by reconstruction are not large.

For illustration purposes, initially  in \S~\ref{sec:results1} we focus only on the high-redshift LRG sample, $0.6<z<1.0$, with a total effective volume of $V_{\rm eff}^{\rm LRG0.7}=0.84\, [{\rm Gpc}\,h^{-1}]^3$ and an effective redshift of $z_{\rm eff}=0.698$, to test the effects of the reference cosmology on the BAO radial and angular distance parameters, $D_H(z)/r_d$ and $D_M(z)/r_d$.  In the baseline flat-$\Lambda$CDM model, the fiducial (catalog) values for the comoving angular diameter distance and the Hubble distance are $D_M^c(z=0.698) = 2576.93\,\rm{Mpc}$ and $D_H^c(z=0.698) = 2984.75\, {\rm Mpc}$, respectively. Along with the fiducial (template) value of $r_d^t=147.879\,{\rm Mpc}$, we obtain $D_M^c(z=0.698)/r_d^t$ = 17.4260 and $D_H^c(z=0.698)/r_d^t$ = 20.1838.

Later in \S~\ref{sec:inference} the results of cosmological inference are presented for the full sample covering the redshift range $0.2<z<2.2$.

\subsection{Cosmological Parameter Inference}
\label{sec:cosmoinference}

We fit power spectrum monopole and quadrupole in the range $0.02<k\,[h{\rm Mpc}^{-1}]<0.30$ with a $\Delta k=0.01\,h{\rm Mpc}^{-1}$, using the model described in \S~\ref{sec:template} to infer the BAO scaling parameters, $\alpha_\parallel$ and $\alpha_\perp$, as defined in Eq.~\ref{eq:scaling}.  The posterior distributions of all the 25 parameters (two scaling parameters of interest and 23 nuisance parameters)  are sampled through a  Monte Carlo Markov Chain (MCMC), using a likelihood ${\Ll}$ given by,  
\begin{equation}
    -2\log {\Ll}({\bf p}|{\bf \Omega}^t,\,{\bf \Omega}^c)=\chi^2 ({\bf p}|{\bf \Omega}^t,\,{\bf \Omega}^c)  = [D({\bf \Omega}^{c})-T({\bf p|\Omega}^{t})] C^{-1}({\bf\Omega}^c) [D({\bf \Omega}^{c})-T({\bf p|\Omega}^{t})]^{T},
\end{equation}
where $D$ denotes the data vector, $T$ the model (theory) vector, ${\bf p}$ the model parameters and  $C$ the covariance matrix (estimated from the mocks). We use uniform improper priors as in table 2 of \cite{gilmarin20}. Given the size of the data vectors (112 as there are 28 $k$-bins per multipole and survey cap), the total number of degrees of freedom is 112-25=87.
In the expression above, we stress the dependence on the cosmologies of the catalog and the template, as described in Fig.\ref{fig:sketch}. Ignoring this dependence in the covariance (which is usually neglected) could yield an underestimation of the errors on the recovered parameters as shown in Appendix~\ref{sec:appB}.

The parameter space is explored through the Metropolis–Hastings algorithm, and the convergence is defined by Gelman-Rubin's convergence diagnostic \cite{Gelman_Rubin}, $R-1<0.005$ computed on 16 independent walkers. 

From the scaling parameters (and their errors) and the adopted catalog and template cosmologies, the BAO distance measures are obtained (Eq.~\ref{eq:scaling}). This provides the posteriors that are used to extract the inferred $D_H(z)/r_d$ and $D_M(z)/r_d$ and their covariance. These measurements can be later used as (compressed) ``data points" to infer the cosmological parameters. In these cases, the model employed is the $k\Lambda$CDM. The cosmological parameters are also obtained via MCMC assuming, as it is customary, Gaussian likelihoods on the BAO distance measurements. The covariance between the BAO distance parameters of the same redshift bin is given by the previous MCMC chain. In the case of non-overlapping redshift bins the covariance across bins is neglected, following the same approach as the official BOSS/eBOSS analyses. The covariance between the first two BOSS LRG bins (which overlap in the region $0.4<z<0.5$) is accounted for through the mocks, by running the BAO analysis in 100 independent realizations. In this case, we have applied uninformative priors on the parameters $\Omega_k$, $\Omega_m$  and $H_0r_d/c$.

\subsection{Methodology and set up of specific tests}
\label{sec:methodology}
Given that the flatness assumption enters separately in the template and in the catalog cosmology (and, of course, the actual realization of the data could correspond to a flat or non-flat Universe) there are several ways in which the flatness assumption can be relaxed which we present below  and will be explored in section \ref{sec:results1}. Variations are with respect to the {\it baseline model} adopted which is a standard flat $\Lambda$CDM with the following parameters:
$H_0= 67.6\, {\rm km\,s}^{-1}\,{\rm Mpc}^{-1}$, $\Omega_bh^2 = 0.021889$, $\Omega_ch^2 = 0.119479$, $\tau = 0.09$, $A_s = 2.040315\times 10^{-9}$, $n_s = 0.97$, and two massless and one massive neutrino species such that $\Omega_\nu h^2 = 0.000644$ and $N_{\rm eff} = 3.044$. The resulting sound horizon scale at radiation drag is $r_d=147.879\,{\rm Mpc}$.

 Since in the standard model $\Omega_m+\Omega_\Lambda=1-\Omega_k$, deviations from flatness can be obtained by varying $\Omega_\Lambda$ or $\Omega_m$. 
We consider both cases: varying curvature through a variation of matter, keeping fixed the dark energy density, $[\Omega_k]_{\Omega_\Lambda=0.69}$, as well as a variation of curvature through the dark energy density, keeping constant the matter density, $[\Omega_k]_{\Omega_m=0.31}$. We study the effect of these changes both in the catalog and in the template cosmologies.
 While in many analyses the catalog and the template cosmology models coincide, in reality, they need not, as schematically represented in Figure~\ref{fig:sketch}. In this figure, green boxes/symbols refer to or indicate the template cosmology dependence, while orange boxes/symbols refer to or indicate the catalog cosmology. Hence, in this paper, we study the following three cases.

\begin{itemize}

\item {\bf Flat catalog cosmology, non-flat template cosmology} to examine the impact of introducing curvature only in the template cosmology, keeping flat the curvature in the catalog cosmology. Keeping fixed $\Omega_m$ (thus varying $\Omega_\Lambda$)  only modifies non-BAO features in the template power spectrum (i.e., the relevant $\pazocal{O}_{\rm lin}$ from Eq.~\ref{eq:Pbao} function is barely modified). On the other hand, keeping $\Omega_\Lambda$ fixed (varying $\Omega_{\rm m}$) does produce strong changes in the BAO oscillatory features as shown in Figure~\ref{fig:olin}.

In Fig. \ref{fig:olin}  the response of the function  $\pazocal{O}_{\rm lin}(k)$   to changes of  $[\Omega_k]^t_{\Omega_\Lambda=0.69}=\pm0.15$ are shown in blue and green respectively,  compared to the fiducial flat case (orange). Both the location and amplitude of the BAO peaks are affected. 

In this work, when $\Omega_k$ is varied by changing  $\Omega_m$, we fix the baryon-to-cold-dark-matter ratio; this keeps the amplitude of the BAO oscillation within a range comparable to that found in observations.\footnote{The alternative would have been to keep $\Omega_b$ constant instead, at the cost of introducing dramatic changes in the BAO amplitude which makes the template unsuitable for data analysis. }

\item {\bf Non-flat catalog cosmology, flat template cosmology} to isolate the effect of curvature in the catalog cosmology. The template cosmology adopted is fixed to the standard baseline flat-$\Lambda$CDM (orange line of Fig.~\ref{fig:olin}). In the calculation of the power spectrum from the galaxy catalog, the conversion between redshifts and distances is done by adopting a non-flat cosmology. We consider both cases, where this departure of flatness is implemented by either keeping fixed $\Omega_\Lambda$ (and varying $\Omega_m$), and the opposite case. Note that in this case, there is no need to impose any restriction on the relative abundance of baryon and cold-dark-matter species, as it has no impact on the $H(z)$ expansion history. Importantly, such non-flat catalog cosmology is consistently applied to both data and mocks, to obtain a covariance matrix which is consistent with the power spectrum of the data. 

\item {\bf Non-flat catalog cosmology, non-flat template cosmology}. 
This is the combined case where flatness is not assumed at any step of the analysis.
As before, we induce this variation by either keeping $\Omega_m$ or $\Omega_\Lambda$ fixed to a value consistent with the standard flat-$\Lambda$CDM model. 

\end{itemize}

\section{Effects on current state-of-the art data: curvature constraints from SDSS BOSS+eBOSS data}
\label{sec:results1}

We first present the effect of dropping the flatness assumption in the template and the catalog separately and only later present the combined effect. As anticipated,  initially we focus only on the high-redshift LRG sample, $0.6<z<1.0$ performing the various test described in \S \ref{sec:methodology}. The motivation for this choice is twofold: the sample is at sufficiently high redshift to maximize the effects of the arbitrary choice of the catalog cosmology, yet it yields a large enough signal-to-noise ratio to appreciate the effects of interest. Later in \S~\ref{sec:inference} the results of cosmological inference are presented for the full sample ($0.2<z<2.2$).

 \subsection{Impact of template cosmology choice}
\label{sec:chtempl}
 \begin{figure}[htb]
     \centering
     
    \subfigure{
     \includegraphics[width=0.479\textwidth]{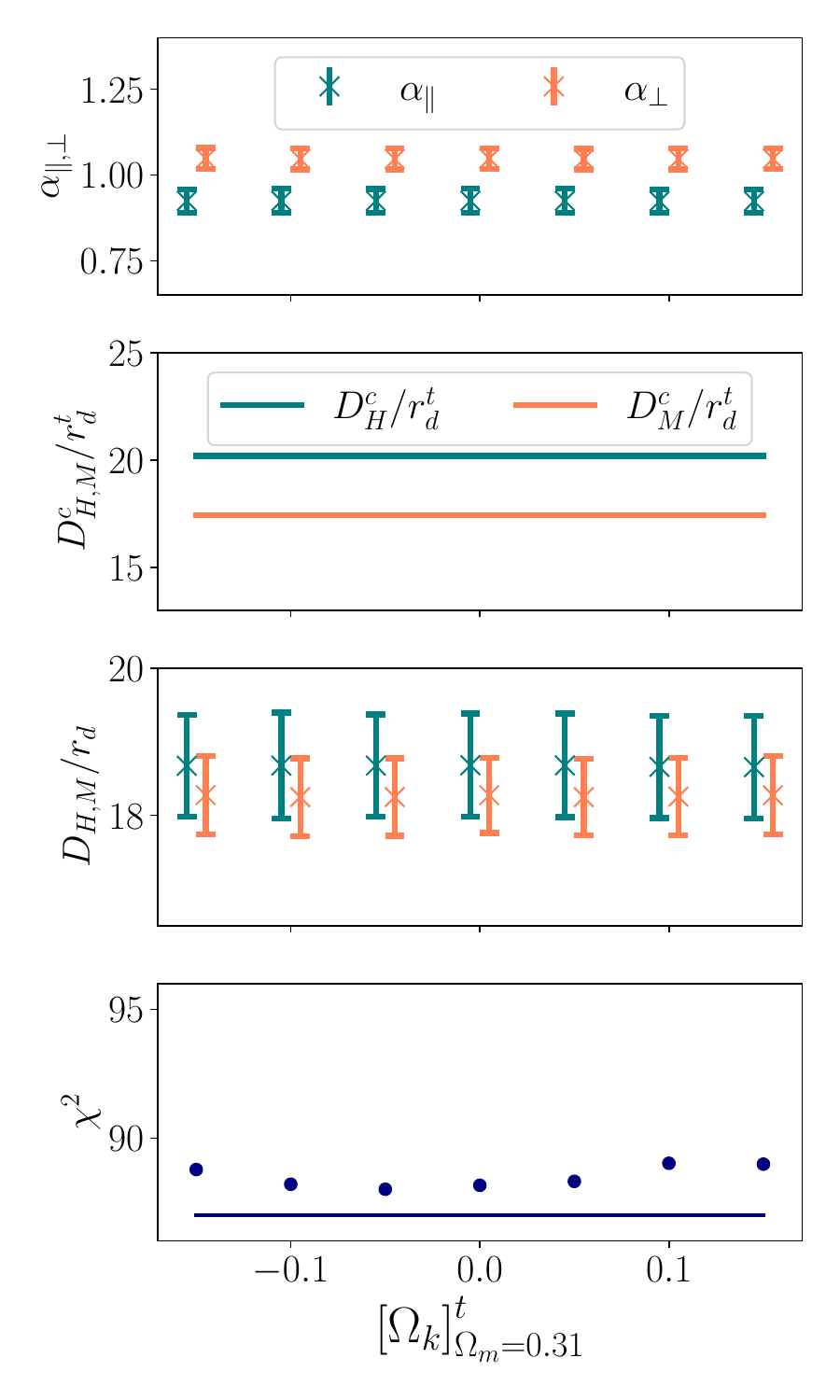}
     }
     \hfill
    \subfigure{
     \includegraphics[width=0.479\textwidth]{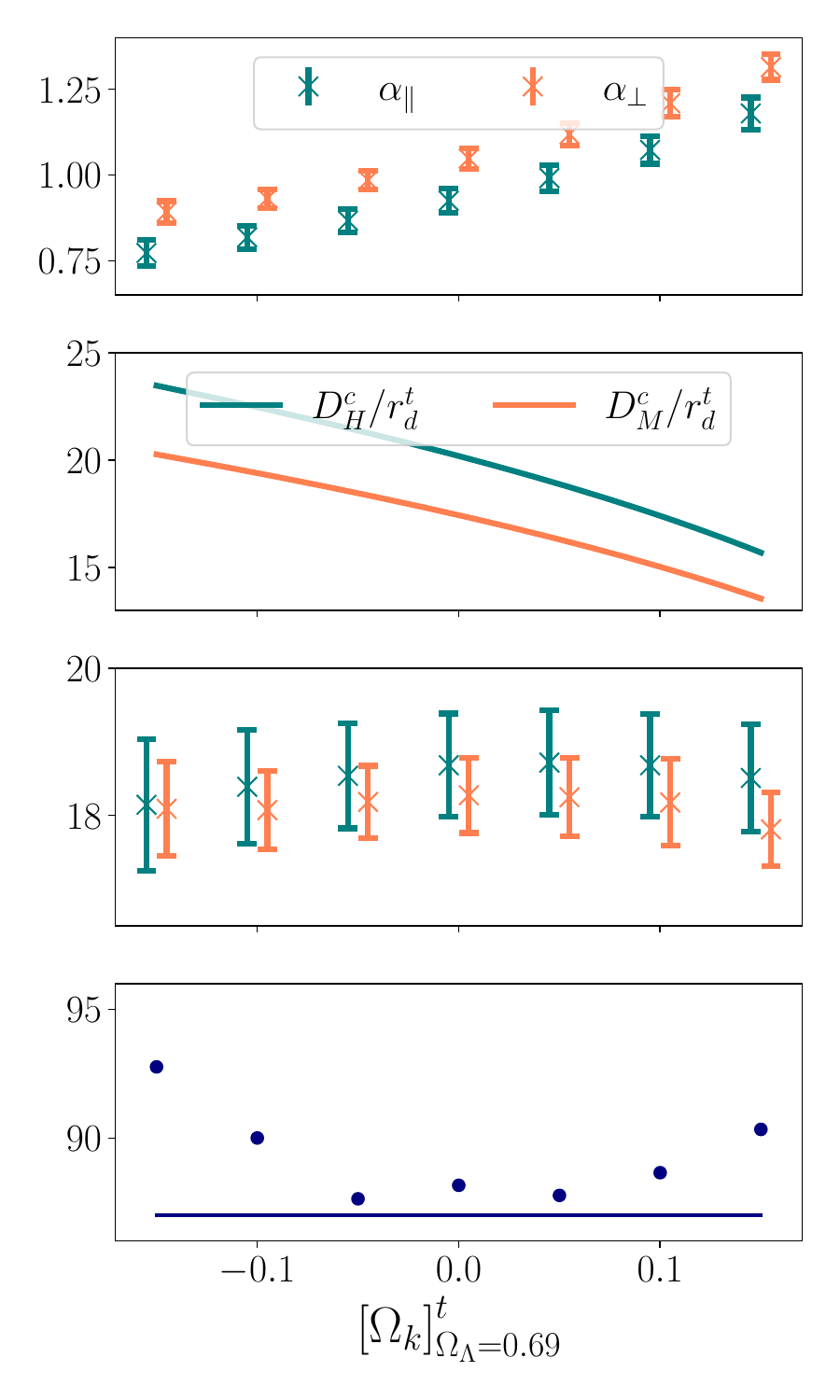}
     }
\caption{Impact of template cosmology curvature on BAO distance measurements from the LRG eBOSS sample ($0.6<z<1.0$, effective volume $V_{\rm eff}^{\rm LRG0.7}=0.84\, [{\rm Gpc}\,h^{-1}]^3$). Curvature is varied through $\Omega_\Lambda$ in the left panels and through $\Omega_{\rm m}$ in the right panels.  For $\Omega_k=0$ the template and catalog cosmologies coincide and coincide with the baseline $\Lambda$CDM model.  First row:  inferred scaling parameters $\alpha_\parallel$ (longitudinal) and $\alpha_\perp$ (transversal) as defined in Eq.~\ref{eq:scaling}.  Second row: predicted BAO distances, $D_H$ (longitudinal) and $D_M$ (angular) in units of the sound horizon scale, $r_d^t$, according to the fiducial template cosmology choice ($x$-axis); Third row: inferred BAO distances corresponding to multiplying the first row by the second row; Fourth row: minimum $\chi^2$ value for the corresponding best fit. Throughout, error bars are 68\% confidence levels. The left panels correspond to variations of the fiducial choice of curvature on the template cosmology (denoted as $\Omega_k^t)$ keeping fixed $\Omega^t_m$ (hence varying $\Omega^t_\Lambda$), while the right panels display variations of the fiducial $\Omega^t_k$ value keeping fixed $\Omega^t_\Lambda$ (hence varying $\Omega_m^t$). The cosmology of the catalog has been left fixed to its baseline (hence flat) model $D_M^c(z=0.698)/r_d^t$ = 17.4260; $D_H^c(z=0.698)/r_d^t$ = 20.1838--, corresponding to the teal and a orange lines on the left panel in the second row--, $\Omega_k^c=0$, $r_d^t=147.879$ Mpc.}     
     \label{fig:ffct} 
\end{figure}

Fig.~\ref{fig:ffct} shows the effect of varying the curvature through $\Omega_\Lambda $ (left panels) and through $\Omega_m$ (right panel) in the template cosmology. These represent the results of the eBOSS LRG data within $0.6<z<1.0$ with an effective redshift, $z_{\rm eff}=0.698$.

Green indicates the line-of-sight quantity and orange is the transversal one.
The first row reports the scaling parameters inferred from the combined Northern and Southern Galactic Caps, 
with the errors derived from the covariance estimated from the 1000 \textsc{EZmock} realizations, as described in \S~\ref{sec:data}.
 The fact that the recovered $\alpha_\parallel$ and $\alpha_\perp$ are not identical to unity indicates that (as expected) the adopted baseline model is not identical to the best fit model preferred by the data, although not at high statistical significance.

The second row shows the predicted distance measures in units of the sound horizon for the assumed fiducial (catalog and template)  models, and in the third row are the corresponding inferred distance measures.  Variations of the template driven by  $\Omega_m$  yield variations in the inferred scale parameters, but the inferred distance measures are stable.  While the recovered distance measures are insensitive to the choice of curvature parameter in the template, not all the templates yield equally good fits to the data, as quantified by the $\chi^2$ in the bottom row.

\subsection{Impact of catalog cosmology, redshift-to-distance transformation choice}\label{sec:z2d}
 \begin{figure}[htb]
     \centering
     
    \subfigure{
     \includegraphics[width=0.479\textwidth]{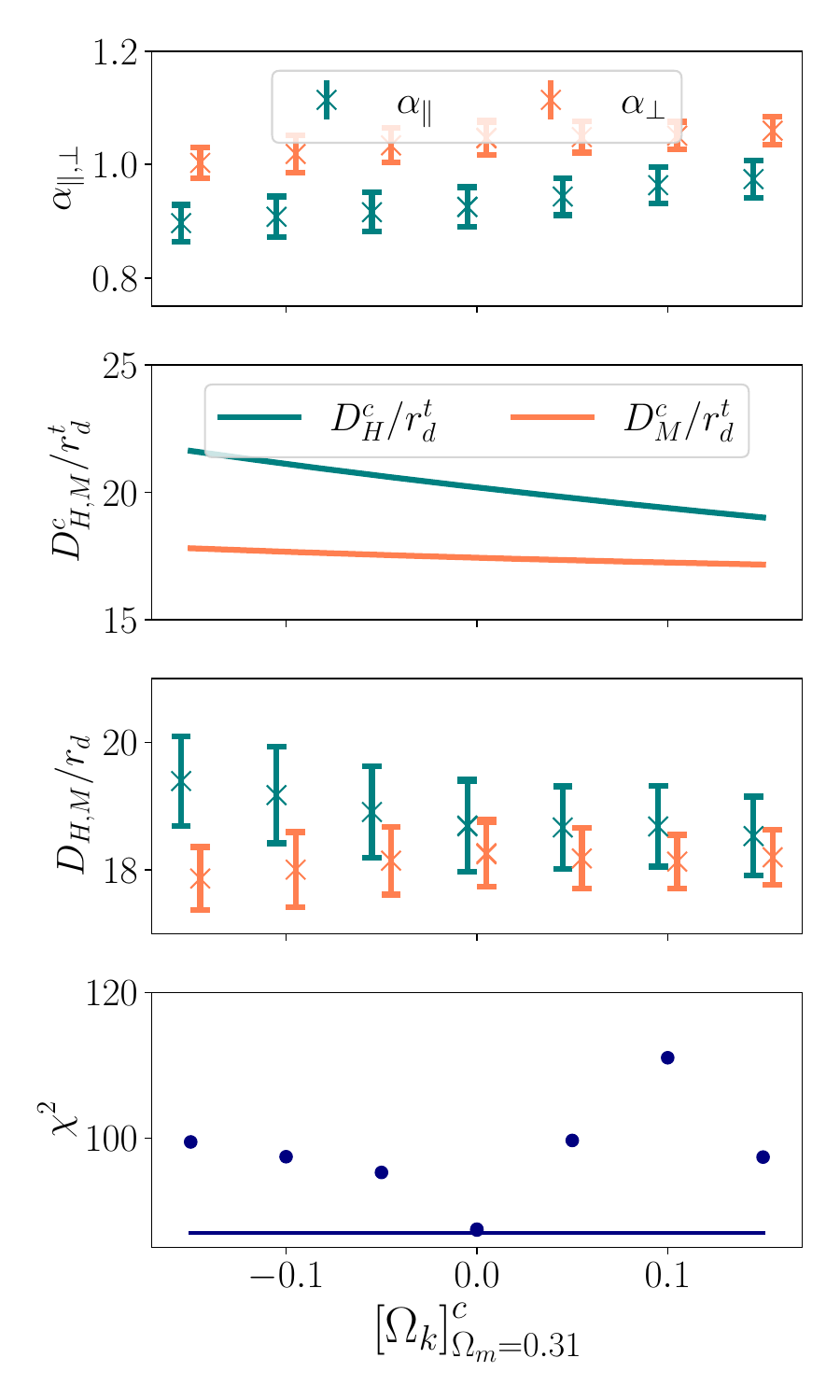}
     }
     \hfill
    \subfigure{
     \includegraphics[width=0.479\textwidth]{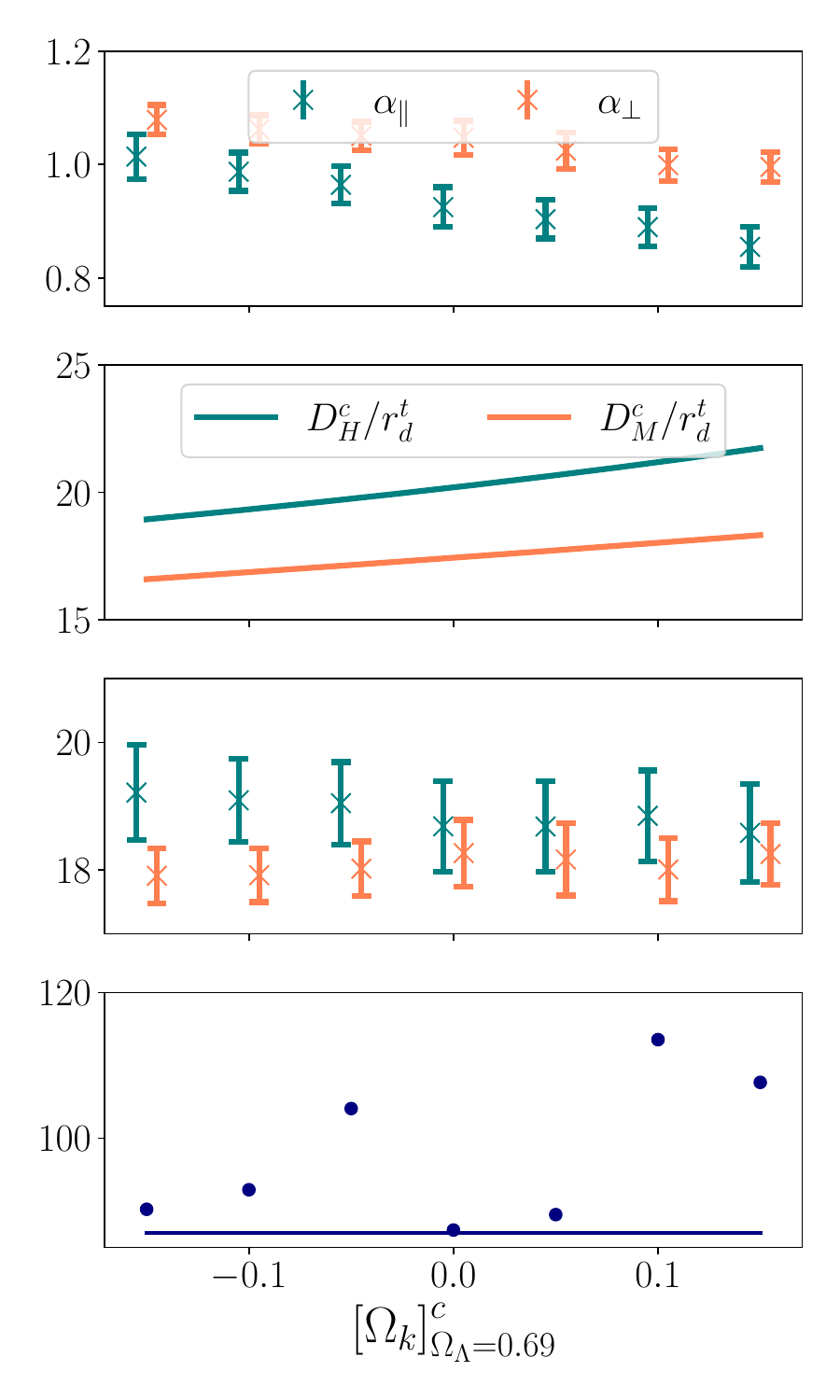}
     }  
\caption{Impact of catalog cosmology curvature on BAO distance measurements.  Measurements are for the eBOSS LRG $0.6<z<1.0$ sample, following the same notation as in Fig.~\ref{fig:ffct}. 
The template cosmology is kept fixed to its fiducial flat baseline model, and the catalog cosmology is changed both for the data and the simulations used to compute the covariance matrix. For $\Omega_k=0$ the template and catalog cosmologies coincide and coincide with the baseline $\Lambda$CDM model.}    
     \label{fig:cfft}
\end{figure}

The choice of the curvature parameter in the fiducial cosmology assumed in the conversion of redshifts into distances (catalog cosmology) affects both the data and the mock galaxy catalogs used to obtain covariance matrices (see Appendix~\ref{sec:appB}). The results are summarized in Fig.~\ref{fig:cfft} where, as in Fig.~\ref{fig:ffct}, in the left column $\Omega_k$  is varied by changing  $\Omega_\Lambda$ and in the right column by changing $\Omega_m$. Green refers to line-of-sight quantities and orange to transverse ones. 

The scaling parameters measured from the eBOSS LRG data (same sample as  in Fig.~\ref{fig:cfft})
are affected by the choice of catalog cosmology, as is the case for the fiducial distance measures.  Interestingly, and differently from the previous case, the dependence of the inferred distance measures on the curvature parameter is weak but noticeable. The minimum $\chi^2$ values for the best fit also degrade much more markedly as $\vert\Omega_k\vert\gtrsim 0.05$ (interestingly enough, there is a region of values around $[\Omega_k]^c_{\Omega_\Lambda=0.69} = -0.15$ in which the goodness of fit appears to be as acceptable as in the flat-$\Lambda$CDM case).
In other words, taking the  $\chi^2$ face value, the eBOSS LRG  data are fully consistent with a flat $\Lambda$CDM model but also appear consistent with $\Omega_{k}\sim -0.15$. For this curvature value, the recovered scaling parameters $\alpha_{\parallel,\perp}$ shown in Figs.~\ref{fig:ffct}-\ref{fig:cfft} are significantly different from unity, which pushes the fixed-template approach is at its limits of validity; we will return to this point in \S~\ref{sec:iterative}.

\subsection{Combined impact of  template and catalog cosmology choices}

 \begin{figure}[htb]
     \centering
     
    \subfigure{
     \includegraphics[width=0.479\textwidth]{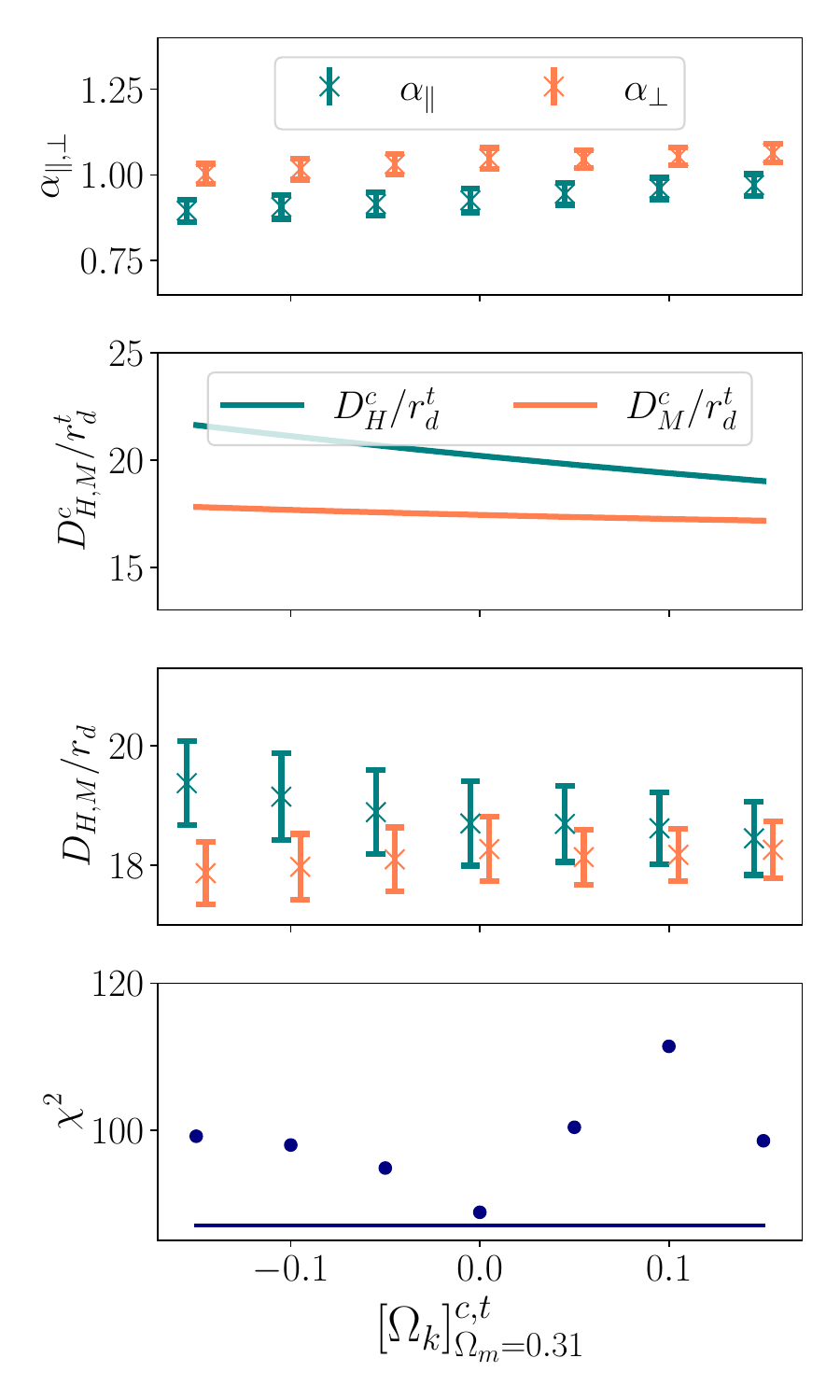}
     }
     \hfill
    \subfigure{
     \includegraphics[width=0.479\textwidth]{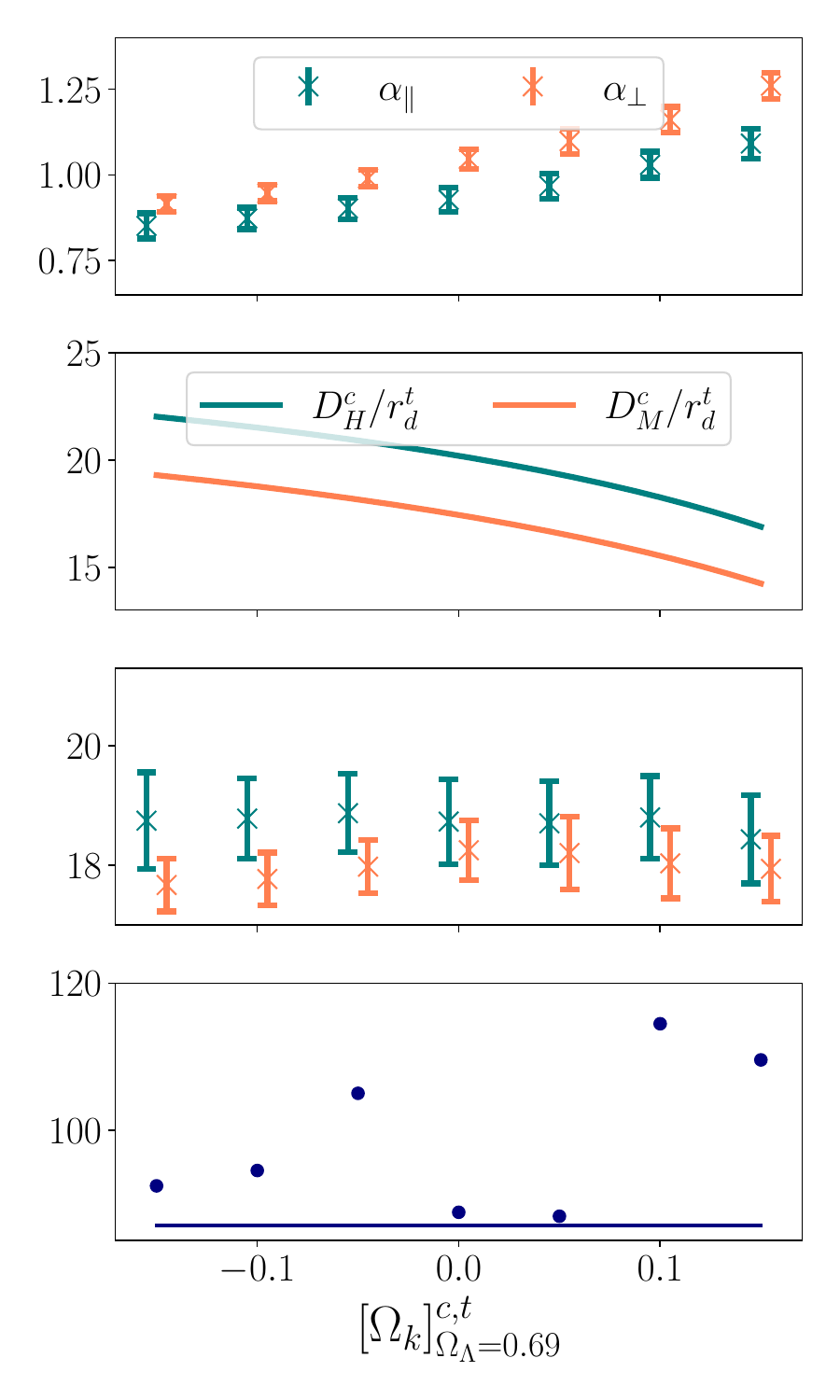}
          } 

\caption{Combined impact of curvature of catalog and template cosmology on  BAO distance measurements for the eBOSS LRG $0.6<z<1.0$ sample, following the same notation as in Fig.~\ref{fig:ffct}. This figure effectively shows 
the combined effect of Figs.~\ref{fig:ffct} and \ref{fig:cfft}. }      
     \label{fig:cfct}
\end{figure}

The results shown in Fig.~\ref{fig:cfct}, which uses the same structure and color scheme as the previous figure, can be readily interpreted in light of the findings of the previous two sub-sections.  

The changes in fiducial values of $D^{c}_{H,M}/r_d^t$ are just a multiplicative result of those reported in Figs.~\ref{fig:ffct}  and \ref{fig:cfft}. Looking at the inferred BAO distances, $D_{H,M}/r_d$ we also can appreciate that the dominant effect is driven by the changes in the catalog cosmology, with the effect of the choice of the template being sub-dominant.

We conclude that extreme shifts in the values of $\Omega_k$ for the catalog cosmology, irrespectively if induced by a change on $\Omega_m$ or $\Omega_\Lambda$, can induce shifts on the BAO distances, which are a fraction of $1\sigma$ for the eBOSS LRG statistical precision. 
On the other hand, inferred BAO distances are quite insensitive to shifts in the template cosmology. 
In particular, we find that for the eBOSS LRGs, varying $\Omega_k$ towards negative curvature values of $-0.15$ in the catalog cosmology produces a shift of up to $1\sigma$ for $D_M/r_d$ and up to $0.5\sigma$ for  $D_H/r_d$. However, these findings do not necessarily imply that the magnitude of these shifts is unchanged when we increase the statistical accuracy of our data (by observing more volume or area). Part of these shifts could be caused by noise in the data, and they do not necessarily represent a coherent shift for independent realizations. To shed light on this issue, in the next section, we explore with more detail this effect focusing on a synthetic dataset much larger (and covering a wider redshift range) than the eBOSS LRG catalog employed here.  

\subsection{Cosmology inference from multiple redshift bins: combined effect of curvature choice in catalog and template cosmology}\label{sec:inference}

We repeat the analysis on the two additional bins of the BOSS LRG sample, $0.2<z<0.5$ with $z_{\rm eff}=0.38$, and $0.4<z<0.6$ with $z_{\rm eff}=0.51$ and on the eBOSS quasar catalog at $0.8<z<2.2$ with $z_{\rm eff}=1.48$. 

For simplicity, we focus only on the case where both the catalog and template cosmologies are varied simultaneously. We also only analyze three fiducial choices, $\Omega^{\rm c,t}_k=\pm 0.15$ and $\Omega_k^{\rm c,t}=0$ under variations of $\Omega_m$ while $\Omega_\Lambda$ is kept fixed at 0.69;  this choice has the largest impact on the oscillatory template as shown in Fig.~\ref{fig:olin}.

Table~\ref{tab:results_data} reports the best-fit parameters of the radial and angular BAO distances in units of the sound horizon scale, $D_H/r_d$ and $D_M/r_d$, respectively, for the four redshift bins covering a total range of $0.2<z<2.2$ for the three  $\Omega_k^{\rm c,t}$ values.

As in the previous sections, we have adjusted the values of the damping BAO parameters (see Eq.~\ref{eq:Pbao}) following the criteria of \cite{seo_eis_2007}: $\Sigma_\perp(z)=10.4D(z)\sigma_8$ $\Sigma_\parallel=(1+f)\Sigma_\perp$. With this, we have determined, $\Sigma_\parallel=9.4\,{\rm Mpc}h^{-1}$, $\Sigma_\perp=4.8\,{\rm Mpc}h^{-1}$ for BOSS and eBOSS LRGs, and $\Sigma_\parallel=8\,{\rm Mpc}h^{-1}$, $\Sigma_\perp=3\,{\rm Mpc}h^{-1}$ for eBOSS QSO, fully consistent with the choices of \cite{neveux20}. There is a small dependence of these parameters with the fiducial cosmology choices, but it is negligible for the precision of BOSS and eBOSS data\footnote{We will relax this assumption when fitting a synthetic data vector with higher statistical power in the next section.}.

\subsection{Implications for constraints on curvature from state-of-the art BAO data}
From the results reported in Table~\ref{tab:results_data} we observe that the inferred values of $D_H/r_d$ and $D_M/r_d$ are very stable even for the extreme variation of $\Omega_k^{\rm c,t}$ considered, and typically fluctuate within a fraction of $1\sigma$. The lower portion of Table~\ref{tab:results_data}  reports the inferred $k\Lambda$CDM parameters from the corresponding  BAO distances. 
 Independently of the chosen fiducial cosmology, the results do not vary significantly, and in all cases the reported $\Omega_k$ value is consistent with 0 within $1.5\sigma$. While the shifts of Table~\ref{tab:results_data} should be revisited in preparation for forthcoming data (see \S~\ref{sec:TNG}), especially for the samples at $z > 0.6$, they do not invalidate the cosmological implications obtained to date in the official analyses of BOSS/eBOSS. 
 What Table~\ref{tab:results_data} shows is that regardless of the fiducial cosmology choice there is no evidence for deviations from flatness from the BOSS/eBOSS data\footnote{Note that in the official SDSS paper \cite{alametal21} also the BAO measurements from Ly-$\alpha$ were added, which since it is a BAO measurement at a higher redshift $z\simeq 2.5$, greatly helps to break degeneracies when combined with the galaxy clustering BAO and improve the constraints on $\Omega_k$. Here we have not used that measurement only because it requires a substantially different pipeline than the one employed to measure the BAO in galaxy clustering catalogs. Note also that the main BAO SDSS results come from analysing the post-reconstructed catalogs. However, here we only refer to those results performed on the pre-reconstructed catalogs, as these are the analyses comparable to the analysis being performed in our paper. The impact of the fiducial cosmology on the reconstruction process goes beyond the scope of this work.}.

Thus, we conclude that even choosing fiducial cosmologies, for both template and catalog, with curvature disfavoured by CMB data (see Fig. \ref{fig:olin}), the retrieved cosmological parameters values are very stable.  For the statistical precision of BOSS and eBOSS data, the choice of cosmology (within the limits explored here) has no significant impact on the conclusion that  BAO data shows a strong preference for a flat universe. These results strongly disfavour claims that the assumption of a flat fiducial cosmology in the BAO analyses biases cosmological inference towards recovering cosmologies which are consistent with a flat universe.

 One might nevetheless worry that the $\sim 1 \sigma$ (statistical) shift induced by different analysis setups being performed on the same data might possibly  indicate a systematic effect which in the near future cannot be ignored as  the statistical errors shrink in future surveys. We argue here that while low-level systematics may be present, they are below the statistical errors and  therefore  do not invalidate the standard results. 
 
In the next section we take a deep dive on  potential  residual systematics: working with a synthetic data vector with higher statistical power and therefore less sensitivity to the intrinsic noise of the data, we identify  and quantify  potential systematic biases (see \S~\ref{sec:TNG}).

\begin{table}[htbp]
    \centering
    \scalebox{0.8}{
    \begin{tabular}{|c|c|c|c|}
    \hline
         SDSS & $\Omega_k^{\rm c,t}=-0.15$ & $\Omega_k^{\rm c,t}=0.00$ & $\Omega_k^{\rm c,t}=+0.15$ \\
        Sample & $D_H(z)/r_d\quad\quad\quad D_M(z)/r_d$ & $D_H(z)/r_d\quad\quad\quad D_M(z)/r_d$ & $D_H(z)/r_d\quad\quad\quad D_M(z)/r_d$ \\
         \hline
         \hline
     LRG $0.2<z<0.5$  & $25.17\pm0.98;\quad10.14\pm 0.23$  & $25.74\pm 0.93;\quad 10.18\pm 0.23$ & $25.67\pm 0.93; \quad 10.29\pm 0.25$ \\
     LRG $0.4<z<0.6$ & $22.23 \pm 1.24; \quad 13.38 \pm 0.29$ & $22.89 \pm 1.27; \quad 13.57 \pm 0.29$ & $23.16\pm 1.30; \quad 13.77 \pm 0.32$ \\
     LRG $0.6<z<1.0$ & $18.39\pm 0.71; \quad 17.87 \pm 0.42$  &  $18.62\pm 0.70; \quad 18.34 \pm 0.53$ & $19.51\pm 0.82; \quad 18.30 \pm 0.53$ \\
     QSO $0.8<z<2.2$ & $13.35 \pm 0.73; \quad 29.53 \pm 0.94$ 
 & $13.32 \pm 0.61; \quad 30.58 \pm 0.98$ & $13.51 \pm 0.63; \quad 31.98 \pm 1.43$ \\
\hline
\hline
$\Omega_k$ & $+0.24\pm 0.23$ & $+0.29\pm 0.23$ & $+0.35\pm 0.26$ \\
$\Omega_m$ & $0.275\pm 0.073$ & $0.239\pm 0.072$ & $0.188\pm 0.079$ \\
$10^2H_0r_d/c$ & $3.255\pm 0.098$ & $3.223\pm 0.091$ & $3.210\pm 0.100$ \\
\hline
    \end{tabular}}
  \caption{BAO distance along ($D_H/r_d$) and across ($D_M/r_d$) the line of sight for different assumed fiducial models, $\Omega_k^{c,t}=-0.15$, $0$, $+0.15$ for the three columns, and for the 4 different BOSS and eBOSS samples with the range $0.2<z<2.2$ which represents a total effective volume of $2.82\,[{\rm Gpc}h^{-1}]^3$ (see \S~\ref{sec:data} for a full description). For each fiducial model, we vary both the catalog and template fiducial cosmologies keeping them equal, and introducing the curvature by varying $\Omega_m$ while keeping fixed $\Omega_\Lambda$, and also keeping the ratio $\Omega_c/\Omega_b$ fixed (see Fig.~\ref{fig:olin} for the different oscillatory patterns as a template fiducial). The last three rows display the inferred cosmological parameter values of a $k\Lambda$CDM model: $\Omega_k$, $\Omega_m$ and $10^2H_0r_d$ (in units of the speed of light, $c$).}
    \label{tab:results_data}
\end{table}
\section{Lessons for ongoing and future surveys: residual systematics and mitigation strategies}
\label{sec:results2}

The next-generation spectroscopy surveys will typically cover an effective volume 10 times larger than that of BOSS and eBOSS, over a comparable redshift range.  Systematics that were safely below the statistical errors for eBOSS (as in previous sections) might become a concern as the statistical precision increases. Here we quantify them and propose who to mitigate them.

\subsection{Systematics quantification and forecasted performance on an effective volume of $28.2\,[{\rm Gpc} h^{-1}]^3$}
\label{sec:TNG}

To this aim, we take our mock data vector to be the power spectrum monopole and quadrupole resulting from the average of 100 individual realizations of the BOSS (\textsc{Patchy}) and eBOSS (\textsc{EZmocks}) samples at the same four redshift bins presented in Table~\ref{tab:results_data}.
We also re-scale the covariance matrix used in \S~\ref{sec:inference} by a factor of 10. By doing so we mimic the statistical precision of a survey covering $0.2<z<2.2$ with an effective volume of $28.2\,[{\rm Gpc}h^{-1}]^3$, but with the intrinsic noise of a survey 100 times larger than BOSS/eBOSS.
Unlike before, while performing the $\alpha_\parallel$ and $\alpha_\perp$ inference,  $\Sigma_\parallel$ and $\Sigma_\perp$ are treated as nuisance parameters to be marginalized over with a uniform prior, and thus having two extra nuisance parameters in our model, the degrees of freedom are $112-27=85$.
As before, the catalog and template cosmologies coincide, and we introduce the curvature variation via $\Omega_m$ variations keeping $\Omega_\Lambda$ fixed. We focus on $\Omega_k=\pm 0.15$ as to maximize the effect on the BAO while adopting a cosmology which, while disfavored, is not fully ruled out by present data. In this scenario, we also aim to determine potential systematic biases in the BAO pipeline in the conditions of (almost) noiseless data vector. In passing and {\it a posteriori} such tests can help partially understand to what extent the shifts observed in the previous section, and shown in Table~\ref{tab:results_data}, are due to noise or an uncorrected systematic component. 

\begin{figure}[htb]
    \centering
    \includegraphics[scale=0.39]{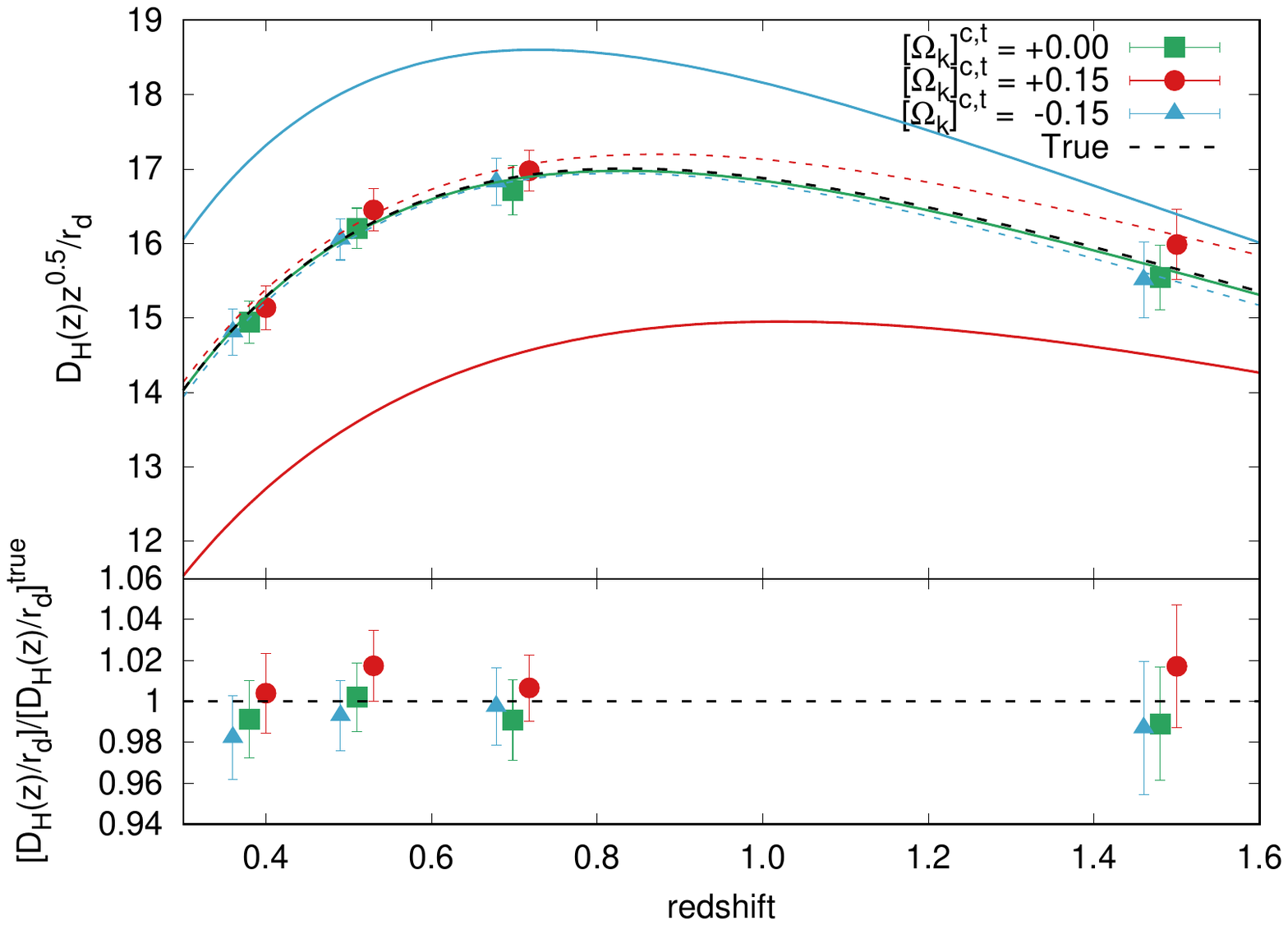}
        \includegraphics[scale=0.39]{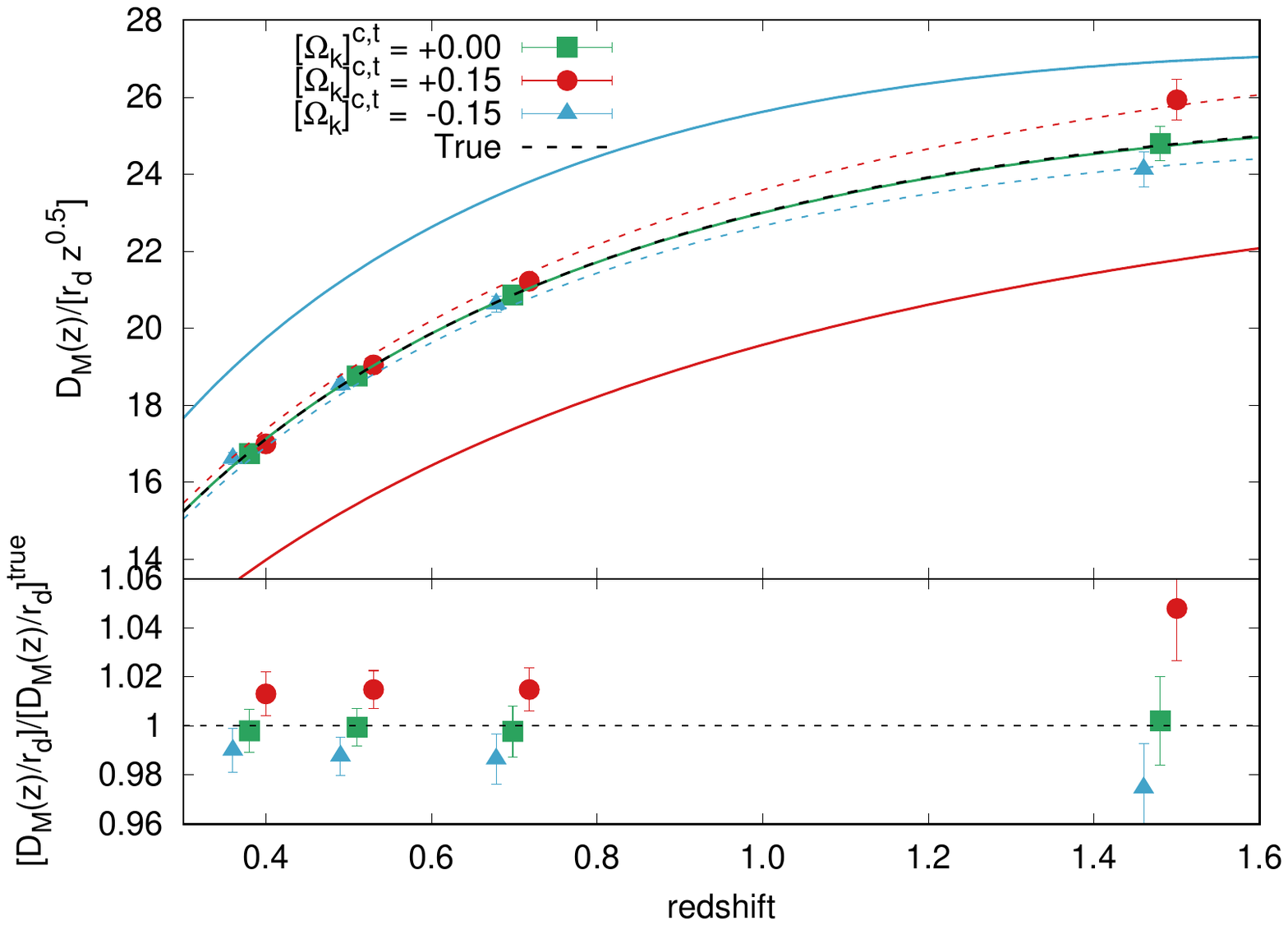}
    \caption{Radial (top) and angular (bottom) BAO distances as a function of redshift. The colored solid lines display the prediction given by the fiducial cosmologies for the values $\Omega_k=0$ (green), $+0.15$ (red) and $-0.15$ (blue), where $\Omega_\Lambda$ is fixed to 0.69. The symbols represent the inferred parameter from a BAO analysis on the average of the power spectra of 100 BOSS/eBOSS mocks with an associated error bar of 10 times the BOSS/eBOSS one, which corresponds to an effective volume of $28.2 [{\rm Gpc}h^{-1}]^3$, mimicking the precision of a DESI-like survey. The color of the symbol references the assumed fiducial cosmology for the BAO analysis: green for $\Omega_k=0$, red for $\Omega_k=+0.15$, blue for $\Omega_k=-0.15$. The red and blue dashed lines display the best-fitting model according to Fig. \ref{fig:inference_av10}. The black dashed line displays the actual cosmology for these mocks, which is very close to the $\Omega_k=0.00$ fiducial choice. The blue and red symbols have been horizontally displaced for clarity.}
    \label{fig:exp_his_10mocks}
\end{figure}

Fig.~\ref{fig:exp_his_10mocks} displays the inferred BAO distances, $D_H/r_d$ (top panels) and $D_M/r_d$ (bottom panels) on this synthetic data-vector.
The color of the symbol indicates the fiducial choice of the template and catalog cosmology as indicated in the legend. The black dashed line corresponds to the underlying true cosmology of the mocks, and the colored solid lines correspond to the prediction for the chosen fiducial cosmology with the same color notation as indicated for the symbols. Note that the flat, baseline, model ($\Omega_m=0.31$ and $\Omega_\Lambda=0.69$) in the green solid line is almost indistinguishable from the true underlying model of the mocks ($\Omega_m=0.307115$, $\Omega_\Lambda=0.692885$)\footnote{See table 1 of \cite{gilmarin20} for a full list of parameters. The true underlying cosmology for \textsc{Patchy} mocks and \textsc{EZmocks} is the same.}.
The recovered distances  $D_H(z)/r_d$, $D_M(z)/r_d$ can be used as observational constraints to infer the parameters of the $k\Lambda$CDM model (see below).  The red and blue dashed lines are the corresponding best-fit distances; in other words, the red (blue) dashed line is the best  $k\Lambda$CDM fit to the red (blue) measurements.
A dashed green line is not displayed as it would be indistinguishable from the solid green and black dashed. 

The lower sub-panels report, for the different assumptions of the fiducial cosmology choice,  the ratio of the measured BAO distances to the expected true value given by the true cosmology of the mocks. The residuals are always within 2\% except for the transversal distance in the QSO case.

The figure shows the systematic bias induced by the choice of the catalog and template fiducial model: positive (negative) curvature fiducial choice yields distances biased low (high). The effect is larger and more statistically significant in the transverse BAO distance $D_M/r_d$ than in the longitudinal BAO distance ($D_H/r_d$). While the systematic trend is evident, as long as $|\Omega_k^{\rm c,t}|<0.15$, the bias on  $D_H/r_d$ is below the statistical error even for a  statistical precision  10 times better than that of BOSS/eBOSS survey. 
The bias in $D_M/r_d$ is particularly important for the quasar sample where it reaches the 2$\sigma$ level and is therefore of some concern.

It is interesting to note that if catalog and template cosmologies have distances significantly over(under) estimated, the recovered distances are biased low (high).  Also, there is a certain asymmetry between the effect of the positive and negative curvature, being the $\Omega_k^{\rm c,t}=+0.15$ case more biased than the $\Omega_k^{\rm c,t}=-0.15$ case. 

Appendix~\ref{app:mask} demonstrates that the systematic shifts on $D_M(z)/r_d$ for the QSO sample reported in  Fig.~\ref{fig:exp_his_10mocks} are mainly driven by the catalog cosmology and not the template cosmology, in agreement with the results of \S~\ref{sec:results2}.

\begin{figure}[thb]
    \centering
    \includegraphics[scale=0.45]{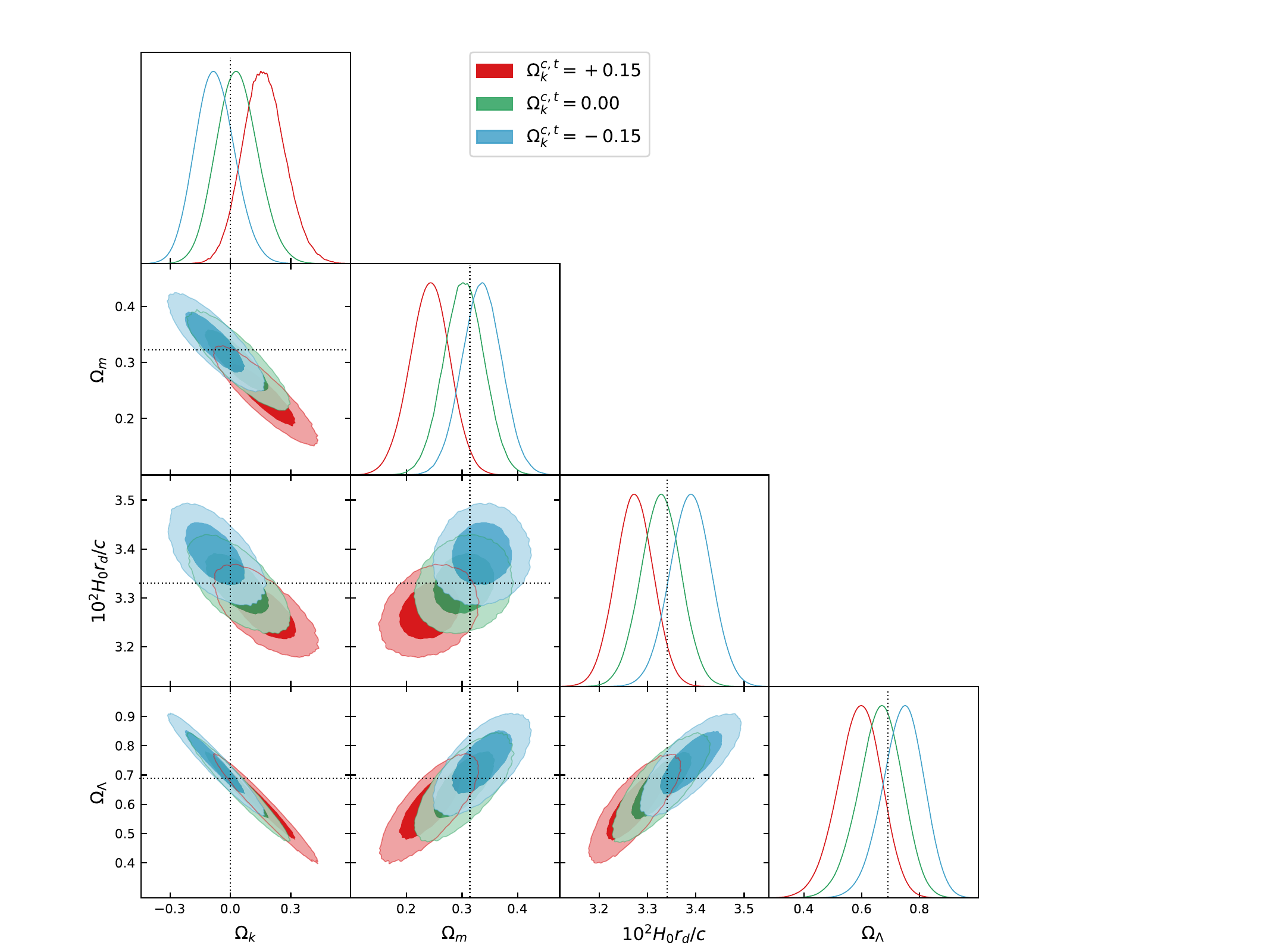}
    \caption{Effect of curvature of the fiducial (template and catalog) cosmology on cosmological parameters inference. These are the posteriors of cosmological parameters of a $k\Lambda$CDM model inferred from the  BAO distance measurements from the mean power spectrum monopole and quadrupole of 100 realizations of \textsc{patchy} mocks and \textsc{ezmocks}, depending on the redshift bin. The 4 redshift bins reported in Table~\ref{tab:results_data} have been used and the employed covariance is scaled
    to mimic a survey with an effective volume 10 times the full BOSS and eBOSS, with $28.2 [{\rm Gpc}h^{-1}]^3$. The color notation follows that of Fig.~\ref{fig:exp_his_10mocks}. The horizontal and vertical black dotted lines display the expected value of the parameter given by the true cosmology of the data vector.    }
    \label{fig:inference_av10}
\end{figure}

Fig.~\ref{fig:inference_av10} illustrates the $k\Lambda$CDM  cosmological parameter posterior inferred from the measured distances, as a function of the curvature in the adopted fiducial (template and catalog) cosmology.
Independent parameters are   $\Omega_k$, $\Omega_m$ and $H_0r_d/c$; $\Omega_\Lambda$ is reported for completeness but is a derived parameter (since $\Omega_\Lambda=1-\Omega_m-\Omega_k$). We choose to report the dimensionless  $H_0r_d/c$ combination, which is most directly constrained by the distance measures (recall that  $r_d$ cannot be constrained from galaxy clustering alone without adopting a  prior on $\omega_b$ and a model for the early Universe).
The dashed red (blue) lines in Fig.~\ref{fig:exp_his_10mocks}  correspond to the best-fit parameters from the red (blue) posteriors.  The corresponding flat case (best fit from green contours) is indistinguishable from the ``true" black dashed line of Fig.~\ref{fig:exp_his_10mocks}.

The systematic shifts induced by the choice of curvature of the fiducial cosmology can be quantified from the results listed in the ``1st iteration" entry of Table~\ref{tab:final_cosmo}, where we report the maximum of the posteriors and the $1\sigma$ confidence intervals (where the first iteration results correspond to the posteriors displayed in Fig.~\ref{fig:inference_av10}). The biased longitudinal and transverse distances reported in Fig.~\ref{fig:exp_his_10mocks} do have a measurable impact in the inferred $k\Lambda$CDM parameters as expected: the fiducial choice does matter.  These shifts are comparable to the statistical errors only for a survey comparable to 10 times the BOSS+eBOSS volume combined, and thus their contribution is completely negligible for the data analysis presented in \S~\ref{sec:inference}. However, for ongoing surveys such as DESI and EUCLID, these shifts must be of concern. In the following section, we show that they arise from a combination of several effects. We also present a proposal of how the analysis should be performed to avoid or minimize such systematic shifts.

\subsection{Systematics and mitigation strategies: The (iterative) solution, the sound horizon scale and geometric effects}
\label{sec:iterative}

The two fiducial cases of cosmologies with $\Omega_k=\pm0.15$ employed in the previous section correspond to cosmologies disfavored by the (synthetic) data vector. The recovered $\alpha_{\parallel,\perp}$ values 
are significantly different from unity (by 10 to 25 \%). In a realistic analysis of survey data, this is a warning flag: the result would immediately raise serious concerns and would not be accepted as a valid and final result. In fact in this case the assumption that  differences in the distance redshift relation between true and fiducial cosmology match a linear scaling which underlie the fixed-template method is expected to break (see e.g.,\cite{Carteretal2020} and references therein). Moreover  the covariance estimate may be inadequate and therefore the error estimate sub-optimal or biased.

\subsubsection{The iterative procedure}
In this section we aim to emulate the (iterative) procedure that should be taken in this case: re-analyze the data using as the new fiducial (or reference),  the best-fitting cosmology of the previous iteration.

In the second iteration step the proposed new fiducial cosmology will still have a curvature different from 0, but should provide a much better fit to the data (in terms of BAO peak-position observable), which in this application corresponds to a standard flat-$\Lambda$CDM. In this fashion, for this second iteration process, the recovered values for $\alpha_{\parallel,\perp}$ are close to unity. This removes the warning flag (as in the first iteration) coming from obtaining a best-fitting $\alpha_{\parallel,\perp}$ different from unity.  However, it is important to bear in mind that even $\alpha \simeq 1$ does not guarantee that the recovered cosmology is correct.

This paradoxical effect (an incorrect cosmology providing best-fitting $\alpha_{\parallel,\perp}$ close to unity) is caused by the (almost) exact degeneracy between the effects in the power spectrum wiggles positions of the catalog and template cosmologies, respectively. The factor  $D_{H,M}/D^c_{H,M}$ (which is different from unity for catalog cosmologies different from the true underlying flat cosmology) is exactly compensated by the factor $r_d^t/r_d$  so that  $\alpha_{\parallel,\perp} = D_{H,M}/D^c_{H,M}\times r_d^t/r_d \simeq 1$, for a set of cosmologies which are different from the true one. Here we are only focusing on the BAO peak position, and not on other BAO features, such as the BAO amplitude, which could raise other warning flags about using a fiducial cosmology which is different from the true one. We will return to these other features at the end of this section. 

\begin{table}[htb]
    \centering
    \begin{tabular}{|c|c||c|c|c|c|c|c|c|}
    \hline
    Name & $\Omega_k^{\rm c,t}$ & $\Omega_m$ & $\Omega_mh^2$ & $\Omega_b$  & $\Omega_k$ & $h$ & $r_d\,[\rm{Mpc}]$ & $r_d'\,[\rm{Mpc}]$ \\
    \hline\hline
       Cosmo$+$ & +0.15 & 0.242 &0.0617 & 0.037228 & 0.167 & 0.505 & 194.34 & 196.91 \\ 
       Cosmo$-$ & $-0.15$ & 0.336 & 0.336 &  $0.052133$ & $-0.077$ & $1.000$ & 100.98 & 100.80 \\ 
        \hline
    \end{tabular}
    \caption{Best-fit cosmologies obtained when fitting the mean of the 100 averaged mocks (with a covariance matrix corresponding to 10 times the volume of a single mock i.e. $V_{\rm eff}=28.2$ [Gpc$h^{-1}$]$^3$ in the baseline model) with the fiducial cases $\Omega^{\rm c,t}_k=\pm0.15$, as displayed in Fig. \ref{fig:exp_his_10mocks} and \ref{fig:inference_av10}. Here  $r_d$ (also noted as $r_d^I$ in Appendix~\ref{sec:appC}) is the sound horizon scale value computed from the sound speed in the early Universe at $z_d<z'<\infty$ (see for eg., Eq.~8 from \citep{ataleofmanyh}). $r_d'$ is 
    the  value of sound horizon length that must be adopted to correct for the mismatch between  $r_d$ and the sound horizon scale as it appears in the (matter) BAO
    (see text for more details). }
    \label{tab:bestfit}
\end{table}

\begin{figure}
    \centering
        \includegraphics[scale=0.39]{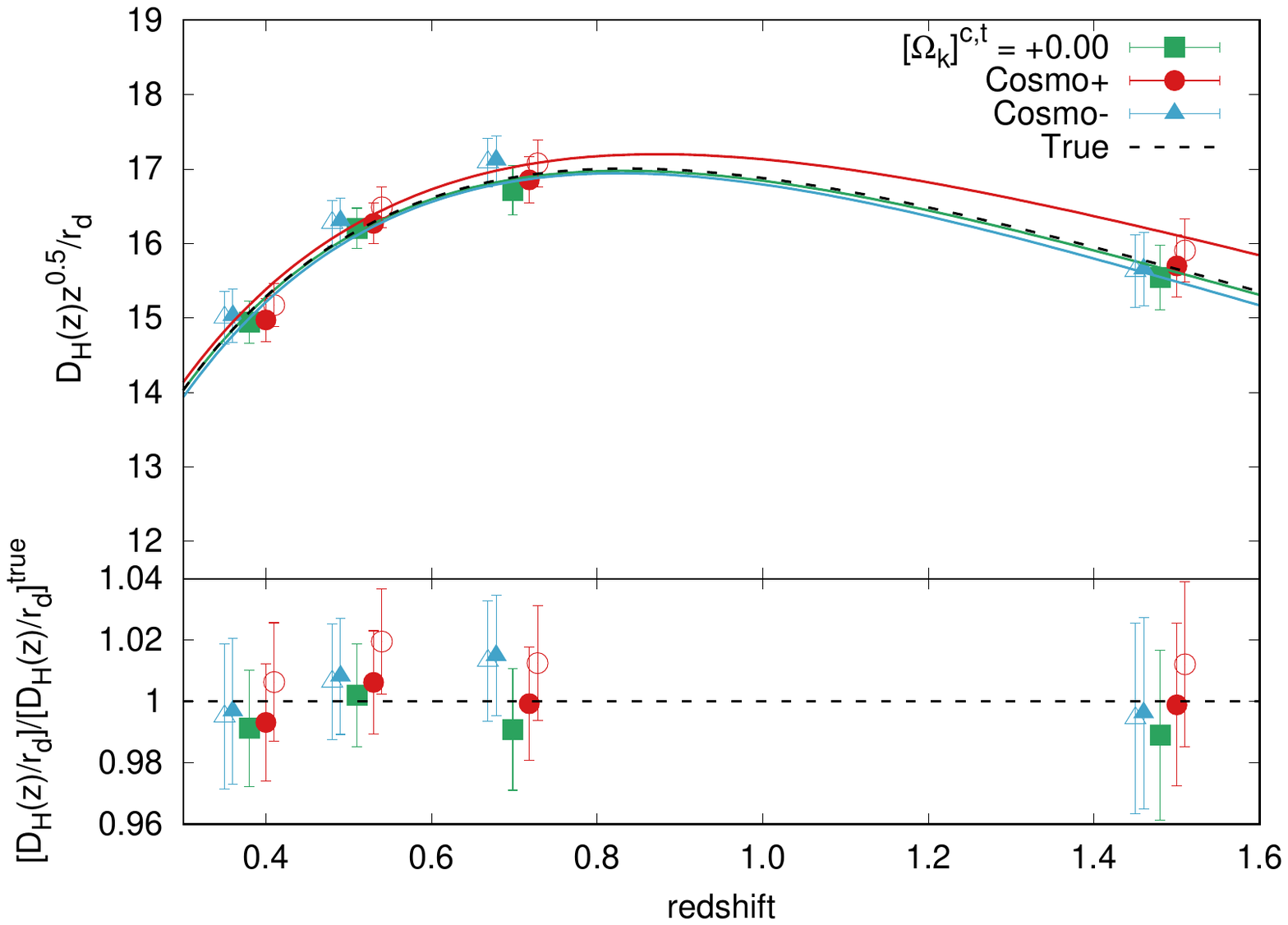}
        \includegraphics[scale=0.39]{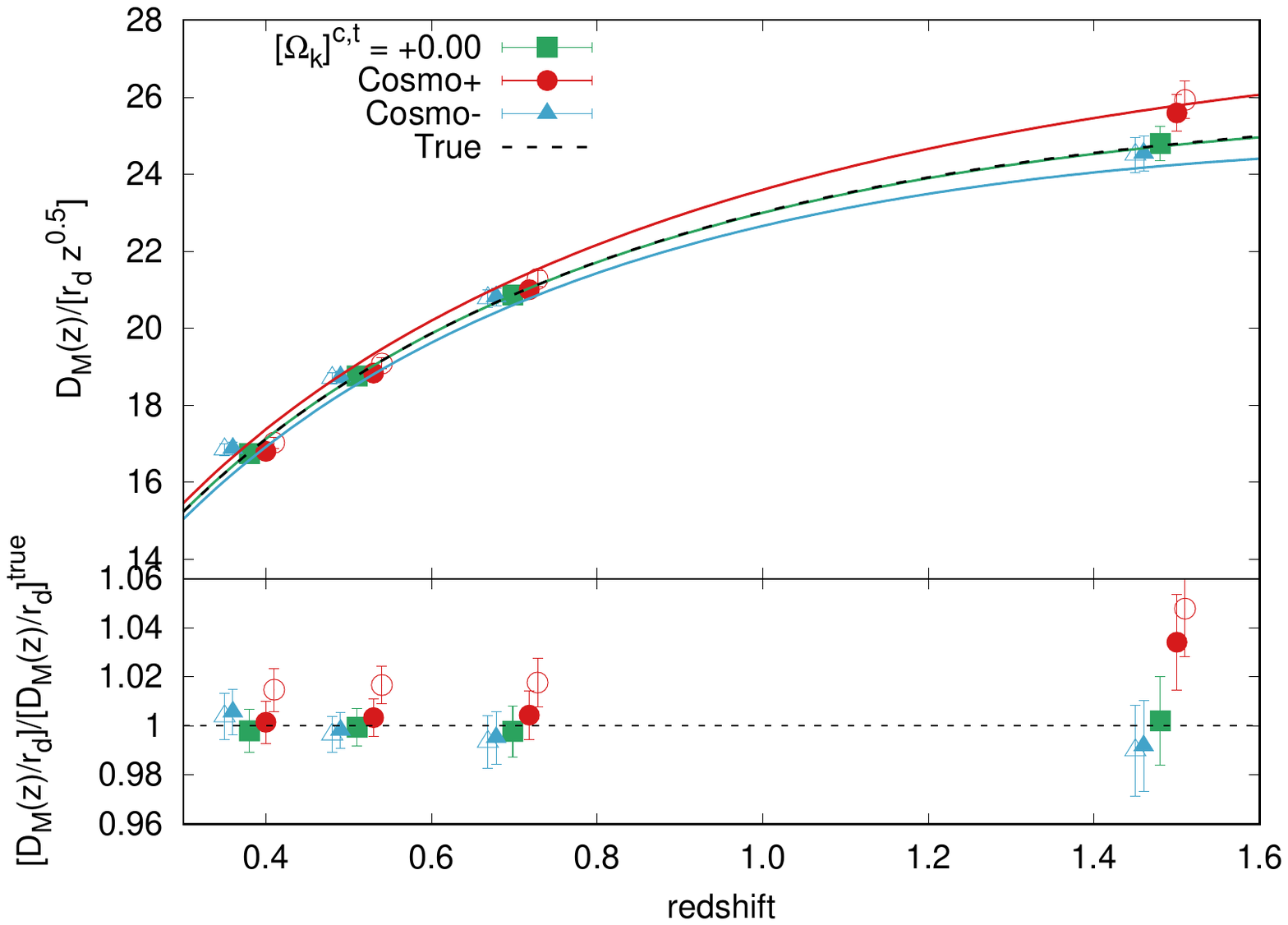}
    \caption{Re-analysis of synthetic data taking as a fiducial cosmology the best-fit cosmologies of the first iteration (displayed in dashed lines in Fig.~\ref{fig:exp_his_10mocks} and as solid lines in the two top panels of this figure). These two new fiducial cosmologies, referred to as Cosmo$\pm$, are also given in Table~\ref{tab:bestfit}. We do not reanalyse the fiducial case of $\Omega_k=0$ as the $\alpha_{\parallel\,\perp}$ were sufficiently close to unity. Consequently, the green symbols are identical to those displayed in Fig.~\ref{fig:exp_his_10mocks}. The empty symbols (red and blue) denote the results of the second iteration analysis without any extra correction. The filled symbols (red and blue) come from substituting the fiducial $r_d$ value obtained from analytically integrating the expansion history up to the drag epoch, by the $r_d'$ proxy given by Eq.~\ref{eq:rdproxy}.}
    \label{fig:reanalysis}
\end{figure}

In our examples, the two new reference cosmologies are listed in Table~\ref{tab:bestfit} and we refer them to Cosmo$\pm$\footnote{Since we do not use any prior on $\Omega_bh^2$ during the cosmological inference we are not sensitive to the baryon density. When we define the fiducial cosmology for the second iteration step, we set $\Omega_b$ by keeping the ratio $\Omega_c/\Omega_b$ constant, as we did in the previous sections.}, depending on the sign of the $\Omega_k$ value. After re-analyzing the data (and the mocks for estimating the covariances) the results are shown as empty symbols (red circles for Cosmo$+$, blue triangles for Cosmo$-$ cosmologies) in Fig.~\ref{fig:reanalysis}. 
In this second iteration the results for Cosmo$-$ shift towards the expected values, now obtaining no significant biases (see empty blue triangles of Fig.~\ref{fig:reanalysis}). On the other hand, for Cosmo$+$ there is a remaining $~2\%$ offset for both $D_H/r_d$ and $D_M/r_d$ (see empty red circles of Fig.~\ref{fig:reanalysis}). Surprisingly, the results for Cosmo$+$ are very similar to those obtained in the first iteration (for $\Omega_k^{\rm c,t}=+0.15$) and displayed as red circles in Fig.~\ref{fig:exp_his_10mocks}.

The iterative adjustment for the fiducial choice of cosmology has solved the mismatch only for the fiducial cosmology with negative curvature but has not fixed the bias in the positive curvature fiducial case.  The remaining $\sim 2\%$ offset in Cosmo$+$ fiducial cosmology is not caused by using fiducial cosmologies yielding $\alpha_{\parallel,\perp}$ values significantly different from unity (in the second iteration $|\alpha_{\parallel,\perp}-1|\lesssim 0.01$, and always deviating with respect to unity at less than $1\sigma$).

The cosmological inference in terms of $\{\Omega_k,\, \Omega_m,\, H_0r_d/c\}$ of this second iteration step is reported in the second row of Table~\ref{tab:final_cosmo}. As expected for Cosmo$-$ we recover a consistent set of parameters with the true underlying cosmology (within $1\sigma$), whereas for Cosmo$+$ we still recover a biased result. 

\subsubsection{The sound horizon at radiation drag  and the BAO scale}

As explained in some more detail in Sch\"oneberg et al. in preparation \cite{Schoneberginprep} and Appendix \ref{sec:appC} and \ref{sec:appD}, this residual $2\%$ bias in the recovered $D_M/r_d$ and $D_H/r_d$, arises from the difference between the sound horizon scale as predicted by early-universe physics (e.g.,  as computed by a Boltzmann code by integrating the sound speed through radiation drag, $r_d^I$)  and the sound horizon length as it appears in the BAO\footnote{The fact that $r_d^I$ may be different from the BAO scale has been discussed for example in \cite{Tseliakhovich_Hirata,Dalaletal2010,Yooetal2011,Slepian_Eis2015}. 
 These authors consider shifts in the BAO scale induced by relative velocities between baryons and dark matter. This effect would be additional to the one considered here, as in our template calculation there is no effect due to relative velocities.} of the linear power spectrum {\it template}, $r_d^T$. 
These two quantities can differ by a few per cent depending on cosmology, an effect already noted by \cite{Lewisrd2014} specifically for cosmologies with $N_{\rm eff}$ variations, but which effectively happens to vary degrees for all model parameters (Sch\"oneberg et al in preparation \cite{Schoneberginprep}). 

The net effect is that, when performing the analysis and cosmological interpretation of BAO with the fixed template approach, the adopted value for the length of the sound horizon at radiation drag needs to be corrected to account for this mismatch i.e., $r_d\longrightarrow r_d'$ (see Appendix~\ref{sec:appC}). For Cosmo$+$ and Cosmo$-$ we determine  $r_d'$ by measuring $r_d^T$ on the oscillatory template of the true, Cosmo$+$ and Cosmo$-$ cosmologies, and using the usual $r_d^I$ values reported by Boltzmann codes. The resulting values are given in the last column of Table~\ref{tab:bestfit}. 
Interestingly, for Cosmo$+$ the relative difference between $r_d$ and $r_d'$ is $1.3\%$, whereas for Cosmo$-$ is $\sim0.1\%$. 

In Fig.~\ref{fig:reanalysis} the filled red and blue symbols are obtained using $r_d'$ for the sound horizon scale. Now all low redshift (galaxy samples) points are consistent with the true values except for the high redshift (QSOs), which remains displaced by about $2\sigma$. We will comment on this later. Also note, that both for Cosmo$+$ and $\Omega_k=+0.15$ the shift on $D_M/r_d$ is not coming from the template; this is shown in Appendix~\ref{app:mask}, where the $\Omega_k=0.15$ is used as a fiducial only for the catalog cosmology (whereas the template is kept to be the true one), and this $\sim2\%$ shift on $D_M/r_d$ is very evident. 

\begin{figure}[thb]
    \centering
    \includegraphics[scale=0.45]{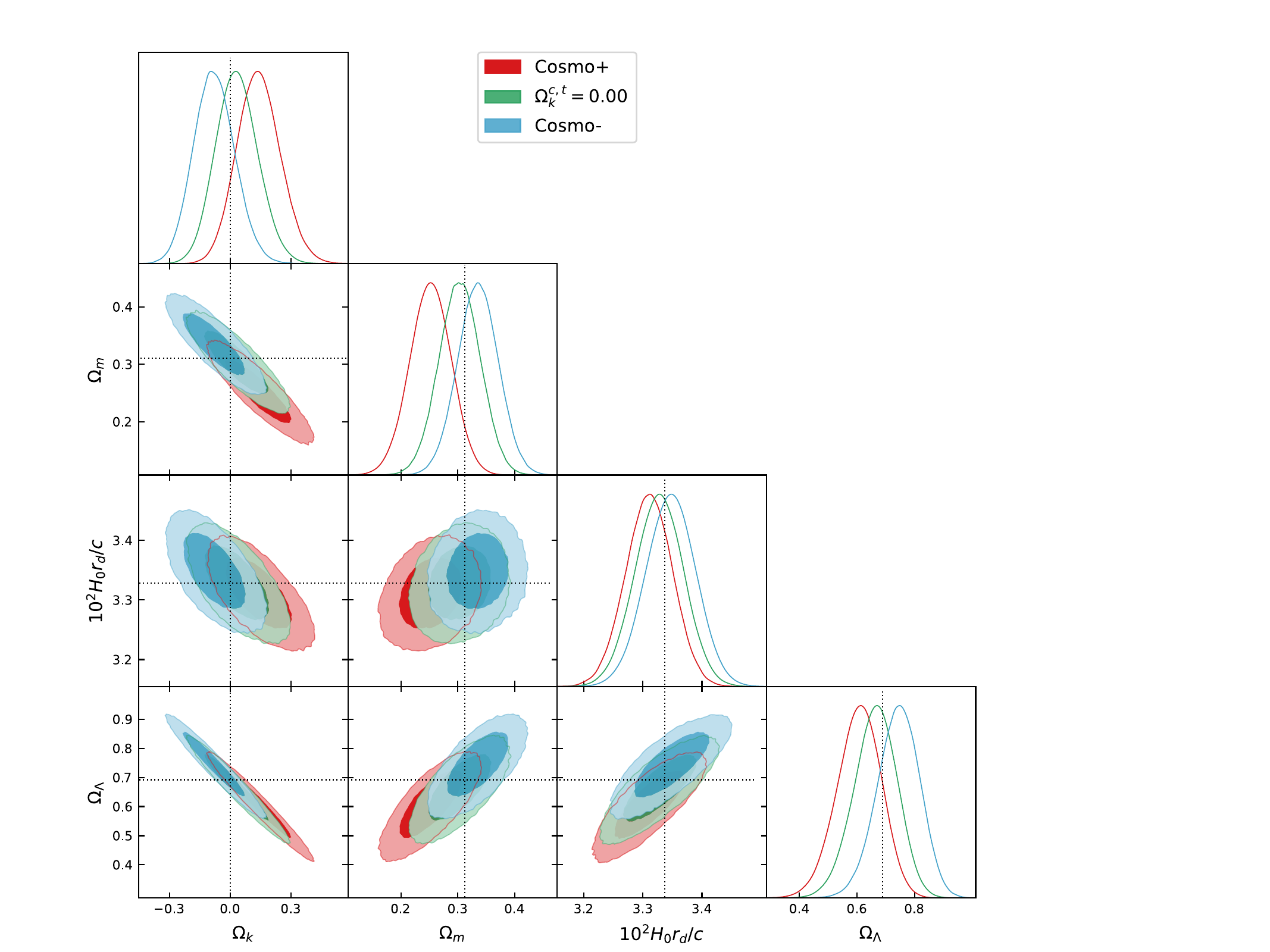}
    \caption{Same as in Fig.~\ref{fig:inference_av10}  but now taking as fiducial cosmology the best-fit cosmologies of the first iteration, namely Cosmo$+$ and Cosmo$-$, reported in Table~\ref{tab:bestfit}, and correcting by the sound-horizon scale $r_d\rightarrow r_d'$ as described by Eq.~\ref{eq:rdproxy} (see text). The green contours are identical to those from Fig.~\ref{fig:inference_av10} as they do not require any correction.
    }
    \label{fig:inference_av10_rdcorr}
\end{figure}

Finally, we display the inferred cosmological parameters after the $r_d$ correction in Fig.~\ref{fig:inference_av10_rdcorr}, as well as in the third row of Table~\ref{tab:final_cosmo}. We see that now the 3 analyses with different fiducial cosmology choices return very consistent values of $H_0r_d/c$ quantity, unlike what happened in our 1st iteration analysis of Fig~\ref{fig:inference_av10}. On the other hand, the deviations concerning the other parameters, $\Omega_k$ and $\Omega_m$ are of around 1$\sigma$ (of the statistical effective volume of $28.2\, [{\rm Gpc}h^{-1}]^3$), similar to those obtained in the 1st iteration run.
In hindsight, this is not too surprising: in this analysis, the constraining power for these parameters is provided by the relative expansion history i.e., by the evolution of $D_M/r_d$ and $D_H/r_d$ with redshift, not by their overall normalization.
We also note that the systematic offsets in $\Omega_m$ and $\Omega_k$ happen precisely in the degeneracy direction, which in a practical case, would be very difficult to disentangle from statistical fluctuations. 

\subsubsection{Geometric effects}
\label{sec:geo}
There is a residual bias in recovered parameters that arises from the redshift dependence of the recovered BAO distances, which is particularly visible in the QSO sample. Such shift is appreciable for this synthetic data set:  it is derived from the average of 100 mocks, but the covariance matrix has been rescaled to correspond to a volume of 10 mocks, hence this analysis is sensitive to systematic shifts as small as 1/3 of a (10 mocks) standard deviation.
Two considerations are in order. 

First, it is well known that non-linearities shift the BAO scale by up to 1\%. Reconstruction alleviates this shift, but here reconstruction is not applied.  As non-linearities are expected to decrease with redshift, the BAO non-linear shift should be less important for QSO than for the low redshift samples, although their higher bias may overcompensate  and induce a relevant BAO shift e.g.,\cite{2402.14070}). Most importantly, the non-linear shift is not expected to differ significantly for the different cosmologies considered. Hence non-linear shift cannot be the dominant source of the residual bias.

Second, the theory and modeling of BAO assume narrow redshift bins, where the bin width is small $\Delta z\ll z_{\rm eff}$ and thus $\Delta D_{H,M}\ll D(z_{\rm eff})$.  This is certainly not the case for the QSO sample.  As the catalog cosmologies vary, so do redshift-distance transformations and the volume element as a function of redshift. For broad redshift bins,   as for the QSO, this introduces geometric corrections of the magnitude seen here (see Appendix~\ref{app:mask} and also Appendix A of \cite{2402.14070}) which are not taken into account. 
In practice, for a survey with a volume more than 10 times that of BOSS-eBOSS, it would not be necessary to consider such broad redshift bins,  the QSO redshift bin would be subdivided, hence mitigating this effect. 
We stress that this residual bias, without mitigation, only becomes appreciable for volumes reaching  $\sim 100$ [Gpc$h^{-1}$]$^3$\,.
Implementing a subdivision of the QSO bin goes beyond the scope of this work given the small size of the effect.

\subsubsection{Performance of the best-fit models}
\begin{figure}
    \centering
  \includegraphics[trim={4cm 300 0cm 10cm},clip=false,scale=0.25]{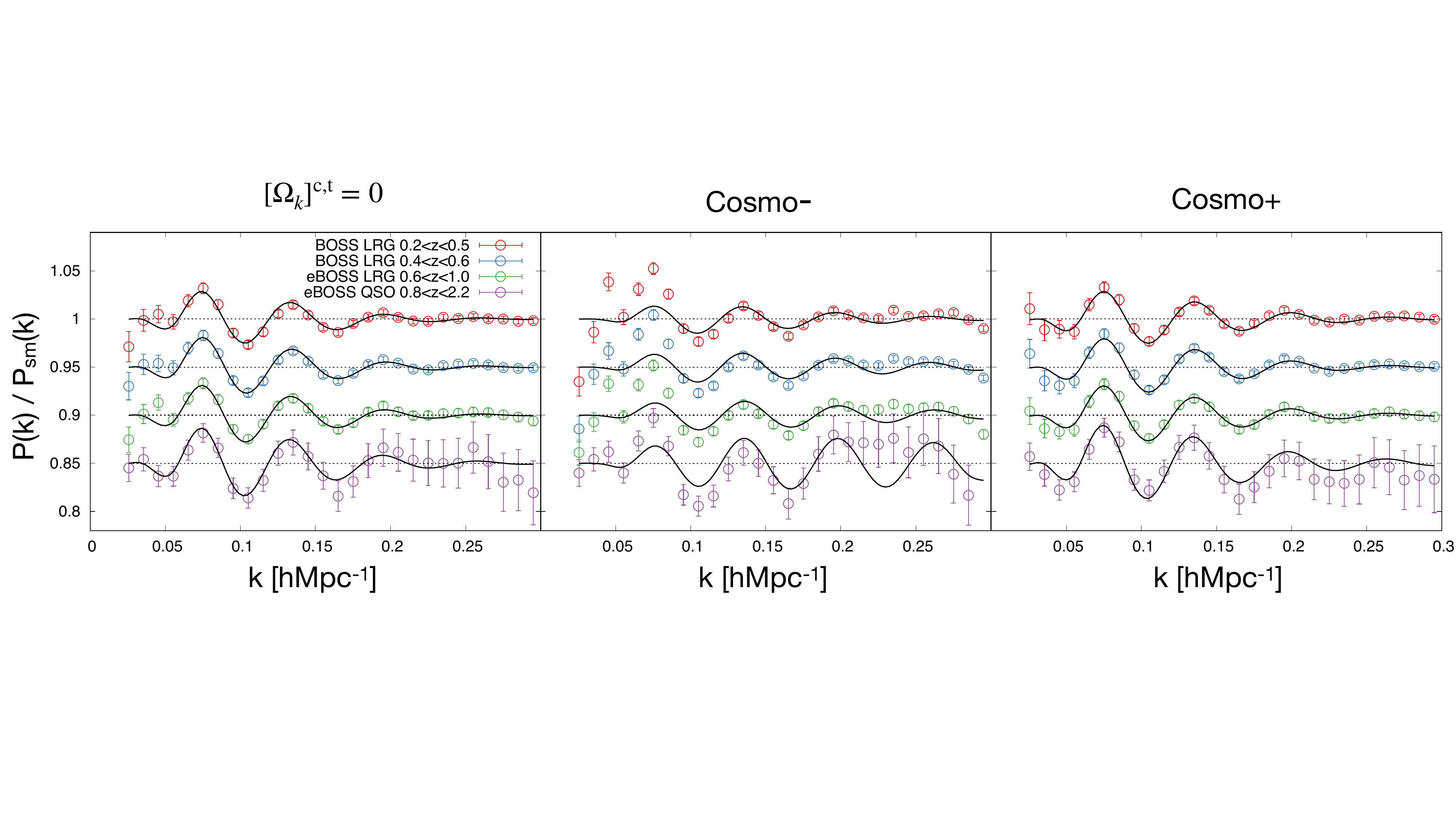}
    \caption{BAO isotropic feature measured in the synthetic data-vector (with the underlying true cosmology flat-$\Lambda$CDM) with a covariance corresponding to a total volume of $28.2\,[{\rm Gpc}h^{-1}]^3$, analyzed according different fiducial (for both template and catalog) cosmologies: flat fiducial cosmology (left panel), Cosmo$-$ fiducial cosmology (middle panel), and Cosmo$+$ fiducial cosmology (right panel) as indicated (see Table~\ref{tab:bestfit} for details). The symbols of different colors represent the different BOSS and eBOSS redshift bins, as labelled. These bins have been displaced vertically for visualization, by 5\%, 10\% and 15\%, as the dotted horizontal lines indicate. The black solid lines display the best-fit model after the 2nd iteration. }
    \label{fig:bao_P0models}
\end{figure}

For completeness, we illustrate the performance of the best-fit model to the oscillatory BAO feature in each of these cosmologies. Fig.~\ref{fig:bao_P0models} displays with symbols the isotropic power spectrum signal measured from the synthetic catalog divided by the smoothed best-fit model when analyzed according to the flat, Cosmo$-$ and Cosmo$+$ reference cosmologies in the left, middle and right panels respectively. The different colors correspond to the 4 redshift bins analyzed: red for LRGs $0.2<z<0.5$, blue for LRGs $0.4<z<0.6$, green for LRGs $0.6<z<1.0$ and purple for QSO $0.8<z<2.2$. The black solid lines correspond to the best-fit model built from the fiducial power spectrum linear template taken from one of these 3 fiducial cosmologies. These best-fit models (and data) have been displaced vertically for visualization purposes by 5\%, 10\% and 15\% as indicated by the horizontal dotted lines.  Unsurprisingly, we observe that for the fiducial flat-cosmology (left panel) the best-fit model reproduces very well the BAO oscillations measured in the catalog. For Cosmo$-$ and Cosmo$+$ the behaviour is unequal. While for Cosmo$+$ the fit can reproduce the BAO oscillatory feature very well, for the Cosmo$-$ the amplitude of the features is not described at the precision required by the data. This unequal behaviour can be traced back to the amplitude of the oscillations in the $\pazocal{O}_{\rm lin}(k)$ fiducial model and their interplay with the nuisance parameters. 
In the Cosmo$+$ model the linear BAO oscillation amplitude in the $\pazocal{O}_{\rm lin}(k)$ is bigger than in the data, whereas in the Cosmo$-$ model, the amplitude is lower than in the data. This can be seen in Fig.~\ref{fig:olins_rs} where the red and blue lines are the Cosmo$+$ and Cosmo$-$ models, respectively, and the green model is for the flat cosmology (which equals to the true cosmology of the data). The higher oscillation amplitude in the Cosmo$+$ model can be corrected by increasing the values of $\Sigma_\parallel$ and $\Sigma_\perp$ parameters in Eq.~\ref{eq:Pbao}, but in the Cosmo$-$ case, the model has no freedom to enhance the oscillations. For this reason, the oscillations on the Cosmo$-$ middle panel have a lower amplitude than required by the data. However, it is remarkable that Cosmo$-$, even providing a bad fit to the full BAO feature, can accurately capture the BAO peak position, in fact sufficiently well to provide unbiased constraints in the cosmological parameters, as reported in Table~\ref{tab:final_cosmo}.

\begin{table}[ht]
\begin{center}
\begin{tabular}{|cc|c|c|c|}
    \hline
           {Starting assumption} &  & $\Omega_k^{\rm c,t}=-0.15$ & $\Omega_k^{\rm c,t}=0.00$ & $\Omega_k^{\rm c,t}=+0.15$  \\
             \hline
             \hline
    \multirow{3}{*}{1st iteration} & $\Omega_k$ &$-0.077\pm0.099$ &$+0.033\pm0.104$ &$0.167 \pm 0.106$ \\&$\Omega_m$ &$0.336\pm0.036$ &$0.303\pm0.036$ &$0.242\pm0.036$ \\& $10^2H_0r_d/c$& $3.389 \pm 0.043$ &$3.327 \pm 0.041$ &$3.272 \pm 0.039$\\
    \hline
    \multirow{3}{*}{2nd iteration}& $\Omega_k$ & $-0.077 \pm 0.102$ &$-$ & $0.141 \pm 0.107$ \\&$\Omega_m$& $0.335 \pm 0.036$ &$-$ & $0.252 \pm 0.036$ \\& $10^2H_0r_d/c$& $3.353 \pm 0.042$ & $-$ & $3.267 \pm 0.039$\\
    \hline
    \multirow{3}{*}{+ $r_d$ correction}& $\Omega_k$ & $-0.078 \pm 0.101$ & $-$ & $0.142 \pm 0.107$ \\&$\Omega_m$& $0.335 \pm 0.036$ & $-$ & $ 0.252 \pm 0.037$ \\& $10^2H_0r_d/c$ &  $3.347 \pm 0.042$ & $-$ & $3.310 \pm 0.040$\\
    \hline
\end{tabular}
\end{center}
\caption{Results from the cosmological inference of the BAO distance parameters $D_H(z)/r_d$ and $D_M(z)/r_d$ for the synthetic data vector with a covariance corresponding to $28.2\, [{\rm Gpc}h^{-1}]^3$ described at the beginning of \S~\ref{sec:TNG}, when analyzed under a $k\Lambda$CDM model (with no priors on the baryon density). The different columns correspond to the different {\it initial} fiducial cosmologies, for both the template and the catalog. The first row matches the results displayed in Fig.~\ref{fig:inference_av10} obtained from the fit to the data assuming the corresponding fiducial cosmology. The second row displays a second-iteration fit, where the data is re-analyzed using as a fiducial cosmology the best-fit cosmology of the previous iteration, namely Cosmo$+$ for the $\Omega_k^{\rm c,t}=+0.15$ column and Cosmo$-$ for the $\Omega_k^{\rm c,t}=-0.15$, both listed in Table~\ref{tab:bestfit}. The third row displays the result of the second iteration but using $r_d'$ instead of $r_d$ as described by Eq.~\ref{eq:rdproxy}, and matches the results of Fig.~\ref{fig:inference_av10_rdcorr}. Note that
no second iteration or  $r_d$ correction apply to the  $\Omega_k^{\rm c,t}=+0.00$ initial fiducial cosmology case.
For completeness, we report the true underlying (expected) quantities of the input cosmology: $\Omega_m=0.307115$, $\Omega_k=0$ and $10^2H_0r_d/c=3.3394$.}
\label{tab:final_cosmo}
\end{table}

We conclude that given the statistical precision of the galaxy and quasar clustering BOSS+eBOSS sample (see Table~\ref{tab:results_data}) the largest observed systematic shifts (those for the fiducial Cosmo$+$) are sub-dominant within the 95\% BOSS+eBOSS confidence level, as summarized in Table~\ref{tab:money-table-sys}. Furthermore, these systematic offsets happen in the degeneracy direction among these 3 parameters.

\begin{table}[]
    \centering
    \begin{tabular}{|c|c|c|}
    \hline
     ${\bf\Omega}$  & $|\Delta {\bf \Omega}^{\rm sys}|$ & $2\sigma_\mathrm{BOSS+eBOSS}$    \\
     \hline
     \hline
       $\Omega_k$ & 0.14 & 0.46 \\
       $\Omega_m$ & 0.055 & 0.14 \\
       $10^2H_0r_d/c$ & 0.029 & 0.18 \\
       \hline
    \end{tabular}
    \caption{Summary of the largest systematic offsets in the $k\Lambda$CDM model observed in the mocks (for the fiducial cosmology Cosmo+ with the $r_d$ correction) along with $2\sigma$ statistical errors associated to the galaxy and quasar clustering BOSS+eBOSS samples.}
    \label{tab:money-table-sys}
\end{table}

\section{Conclusions}\label{sec:conclusions}

The impact of the assumption of a non-flat fiducial cosmology on the
measurement, analysis and interpretation of BAO distance variables, along and across the
line-of-sight has been misunderstood and overlooked.  In this paper, we have clarified and quantified the relevant effects.  The main takeaway points are as follows.

 The assumptions about cosmology enter in the transformation of tracer’s redshifts into distances (the catalog cosmology) and, for the so-called fixed-template analyses,  on the choice of the base template (template cosmology).

 For curvature, the dependence of the inferred distance measures on the catalog cosmology is much larger than the dependence on the template cosmology.  For present-day surveys (i.e., BOSS+eBOSS), the choice of $\Omega_k$ may introduce up to $\sim 2\sigma$  shifts on the inferred cosmological parameters  
but only when the fiducial $\Omega_k$ in the catalog cosmology is significantly (unrealistically) different from the true one.

When considering a  synthetic data vector subject to much lower cosmic variance noise, the corresponding systematic shifts can be quantified better and (in terms of the galaxy and quasar clustering of BOSS+eBOSS data) become up to $\sim0.5\sigma$.

While the recovered distance measures are insensitive to the choice of curvature parameter in the template, not all the templates yield equally good fits to the data. A simple beyond-BAO-only analysis can spot that (like in Fig.~\ref{fig:bao_P0models} for the Cosmo$-$ model). 

 It is important to note that since the most important effect of the choice of fiducial cosmology is in the catalog,  full modeling approaches (as they are implemented now) are not exempt from these potential systematic effects. In specific applications where full modeling approaches achieve error bars on the inferred cosmological parameters smaller than those obtained with template-based approaches (see e.g.,~\cite{brieden_shapefit,brieden_ptchallenge} and refs therein), this should be an important concern.

 Unsurprisingly, the choice of the template cosmology matters when the chosen baryon fraction $\Omega_b/\Omega_m$, produces BAO oscillations of an amplitude that does not match the signal in the data. However, the standard pipelines already ensure that the BAO amplitude in the template is well-matched to that of the data, keeping in mind that in the modeling, nuisance parameters can suppress the BAO signal but cannot enhance it. This could be fixed with some simple modification of the pipeline.

The effects of the catalog cosmology are important not just on the signal (i.e., in generating the catalog) but also in the estimation of the covariance matrix for the cosmological analysis in those cases this covariance is inferred from mocks. This is especially crucial when the covariance matrix is estimated from mock catalogs. In principle, the mock catalogs need to be transformed using the same catalog cosmology as the data. Ignoring this leads to mis-estimate error bars on cosmological relevant quantities up to a factor of 2 in some extreme cases.
In practice, we demonstrate that there is a simple and computationally very cheap way to account for this which involves simply suitably rescaling the covariance computed at a fiducial cosmology.

 Despite all this, the impact on cosmological inference of all the above effects (dominated by the choice of the catalog cosmology), at present is not large enough (compared to the statistical errors) to be a significant worry and does not invalidate current constraints from BOSS+eBOSS data (see Tables \ref{tab:results_data} and \ref{tab:money-table-sys}). The current BAO constraints are consistent with flat geometry and do not show indications for any deviation from flatness (some recent fearmongering e.g., ~\cite{diValentinocurvature2021} is unfounded). Moreover, the systematic effects discussed here can only affect the inference of $H_0$ by no more than $\sim 2$\%, and only in some extreme cases which are highly disfavored by other probes. Hence this effect should not be invoked to resolve the current Hubble tension (see e.g., \cite{ataleofmanyh} and reference therein).

 For future surveys, however, given the smaller statistical errors, the choice of the fiducial (catalog) cosmology matters: some choices may introduce systematic shifts that are statistically significant. 

 There is an age-old guideline, which is implemented e.g., by BOSS/eBOSS official BAO pipelines: if the inferred $\alpha_{\parallel, \perp}$ differ from unity by more than a few \%,  the template based analysis stops being reliable, and results should not be trusted. A better template must be used.
This may be promoted to a  golden rule: if the recovered  $\alpha_{\parallel, \perp}$ differ from unity by more than few \%, one must adopt an iterative approach whereby not only the template is corrected but, importantly,  the catalog is regenerated using the best-fit cosmology, and the analysis repeated. This is expected to converge rapidly (at the 2nd iteration). 

 In a practical application, even when adopting a full modeling approach, there is value in doing a fast (template-based) iterative analysis first. A full analysis can proceed after the catalog has been regenerated to yield $\alpha_{\parallel,\perp}$ close to 1.

 Some of the systematic shifts presented here are due to the (almost) exact degeneracy between the effects in the power spectrum wiggles position of the catalog and template cosmologies i.e.,  a change in the catalog cosmology and a change in the template cosmology combine so that the resulting change in $D_{H,M}/D^c_{H,M}$  is exactly compensated by a change in $r_d^t/r_d$ (hence $\alpha_{\parallel,\perp}$ are unaffected). The specific direction (and possibly the extent) of the resulting degeneracy in the cosmological parameters space may depend on the survey and on how the $z-$bins are chosen.

 For the level of precision of forthcoming experiments, the (small and to-date neglected) mismatch between $r_d^I$ (the early-universe sound horizon at radiation drag) and $r_d^T$ (the sound horizon scale as it appears in the BAO) may not be negligible anymore. Ignoring the effect at the inference level might bias the recovered cosmological parameters, especially $H_0$ at the 1-2\% level at the maximum. The correction can be computed but it is time-consuming: further studies on this topic will be presented in \cite{Schoneberginprep}.

 In this paper we have focused on extensions of standard $\Lambda$CDM involving curvature: it is reasonable to expect that similar effects may happen with other model extensions that affect the redshift-distance relation such as the so-called $w$CDM or $w_0w_a$CDM models.

Before we conclude the reader may wonder what the effects may be for blind analyses of spectroscopic surveys such as the one proposed by \cite{BriedenBlinding}. Such blind analyses rely on introducing (blind) changes in the redshift-to-distance mapping. Since the changes are small and the full analysis is re-done on the catalog after unblinding all systematic shifts (except the one introduced by the mismatch of $r_d^I\ne r_d^T$) are corrected or are subdominant compared to the statistical errors for surveys up to $\sim 50$ [Gpc/$h$]$^3$.

In summary,  at present, the fiducial cosmology assumptions built-in in current analyses do not bias any significant results from publicly available surveys.
We envision that the effects uncovered here and the mitigation measures proposed will be useful in improving the robustness of the analysis and interpretation of forthcoming surveys.

\acknowledgments
We thank Nils Schoneberg for key suggestions in the final stages of the paper and invaluable help with the computation of the `template sound horizon scale values'. 
SSW acknowledges funding from Plan Propio de Investigaci\'on de la Universidad de C\'ordoba (2022) through Submodalidad 2.1 (Becas ``Semillero de Investigaci\'on'').  HGM acknowledges support through the program Ramón y Cajal (RYC-2021-034104) of the Spanish Ministry of Science and Innovation. AJC acknowledges support from the Spanish Ministry of Science and Innovation (PID2022-140440NB-C21), and from the European Union - NextGenerationEU and the Ministry of Universities of Spain through \textit{Plan de Recuperaci\'on, Transformaci\'on y Resiliencia}. AJC thanks the hospitality of Departamento de F{\'\i}sica Te\'orica y del Cosmos at Universidad de Granada. LV and HGM acknowledge the support of “Center of Excellence Maria de Maeztu 2020-2023” award to the ICCUB (CEX2019-000918-M funded by MCIN/AEI/10.13039/501100011033) and project PID2022-141125NB-I00 MCIN/AEI.

\appendix

\section{Distances in curved space}
\label{sec:appendixA}
For completeness, we report here key equations for the Hubble function $H(z)$, the Hubble distance $D_H(z)$, the comoving distance $D_C(z)$, the comoving angular diameter distance $D_M(z)$  and the angular diameter distance $D_A(z)$ for a curved universe \cite{hogg},
\begin{align}
H(z) &= H_0 E(z) \\
D_H(z)  &= \frac{c}{H(z)} \\
    D_C(z) &= \frac{c}{H_0}\int_{0}^{z} \frac{dz'}{E(z')} \equiv \frac{c}{H_0}I(z) \\
    D_M(z) &= \begin{cases}
\frac{c}{H_0 \sqrt{\Omega_k} }\sinh [ \sqrt{\Omega_k} I(z) ]                                             &\Omega_k>0 \\
D_C(z)& \Omega_k =  0 \\
\frac{c}{ H_0\sqrt{|\Omega_k|}} \sin [ \sqrt{|\Omega_k|} I(z) ]                                                  &\Omega_k <0\\
\end{cases}\\
D_A(z)&=\frac{D_M(z)}{1+z}
\end{align}
where $H_0 = H(z=0) = 100h$ km s$^{-1}$ Mpc$^{-1}$, and $E(z) = \sqrt{\Omega_m(1+z)^3+\Omega_k(1+z)^2+\Omega_\Lambda}$.

\section{Fiducial cosmology effect in the covariance and rescaling prescription}
\label{sec:appB}

In this paper, we consistently re-compute the particle positions and the power spectrum of the mock catalogs as the catalog cosmology is changed. This is the computationally expensive approach to obtain the correct covariance matrix. This appendix demonstrates that there is a cheap and efficient alternative which consists of computing the covariance from the mocks for a fixed fiducial cosmology and then applying the rescaling given by the isotropic correction factor.

 As discussed in \S~\ref{sec:z2d}, when analyzing the data catalogs an arbitrary (fiducial) cosmology is adopted for transforming redshift into distances. This arbitrary set of parameters $\{\Omega^{\rm c,t}_m,\,\Omega^{\rm c,t}_\Lambda\}$ has to be consistently applied to the computation of the power spectrum of the data and of the mock catalogs used to estimate the covariance matrix of the data. Keeping the covariance matrix fiducial cosmology unchanged while varying the fiducial catalog cosmology for the data induces an under or over-estimation of the errors.

The physical reason behind this effect is that when the fiducial cosmology for the redshift-to-distance transformation on the data is changed,  the level of the sparseness of the data catalog changes, and an anisotropic signal is potentially introduced. If this is not reflected in the mocks used to estimate the covariance, the resulting covariance will not represent accurately the errors of the data. 

Since this correction is potentially a small effect (after all it is an error on the error), we quantify it here for the variations of fiducial cosmology studied in this paper. 

\begin{figure}[htb]
    \centering
    \includegraphics[scale=0.4]{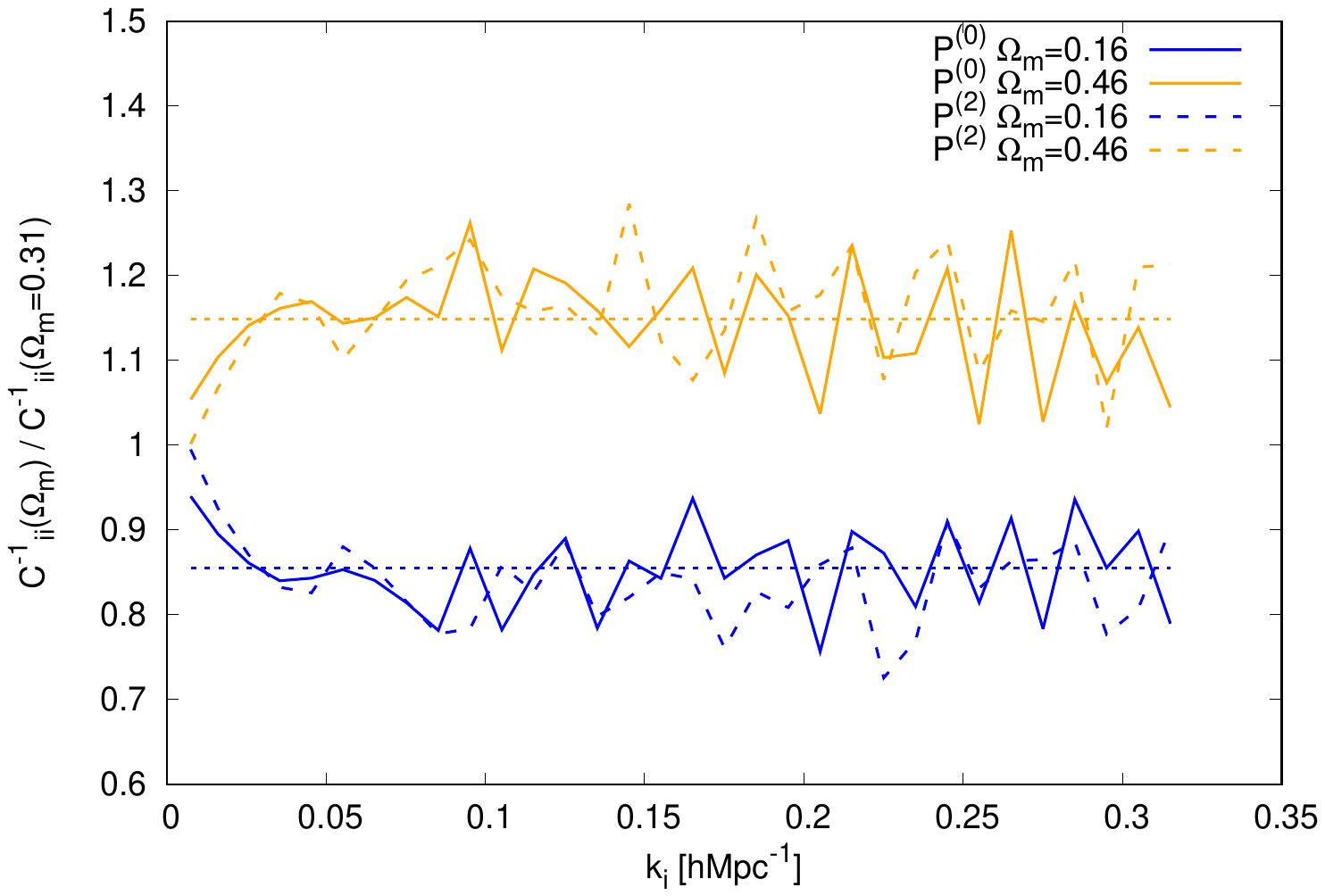}
    \caption{Scaling factors for the inverse of the covariance matrix derived from mocks (NGC eBOSS LRG at $z_{\rm eff}=0.698$) for the monopole (solid lines) and quadrupole (dashed lines) diagonal elements for $\Omega_m=0.16$ (blue) and $\Omega_m=0.46$ (orange), relative to the $\Omega_m=0.31$ reference case; $\Omega_\Lambda$ is fixed to 0.69. In all cases, the covariance has been estimated from 1000 \textsc{EZmock} realizations. For reference, the analytical isotropic factors (derived in Table \ref{tab:scaling_cov_factors}) are represented in dotted lines.}
    \label{fig:rescaling_comparison}
\end{figure}

Fig.~\ref{fig:rescaling_comparison} shows the ratio of the inverse of the diagonal covariance elements ($C^{-1}_{ii}$ are the square of the error associated with the $i$-th bin power spectrum, $P(k_i)$) inferred from the mocks for the cases $\{\Omega_m,\,\Omega_\Lambda\}=\{0.16,0.69\}$ (blue lines), and $\{\Omega_m,\,\Omega_\Lambda\}=\{0.46,0.69\}$
(orange lines), respect to the baseline $\{\Omega_m,\,\Omega_\Lambda\}=\{0.31,0.69\}$ case. The solid lines represent the elements of the block-diagonal case for the monopole, and dashed lines for the quadrupole. We report that deviating from the baseline $\{0.31,0.69\}$ systematically over- or under-estimate the errors of $P^{(\ell)}$ by factors of $\sim 7\%$ ($\sim15\%$ on the errors squared), on both monopole and quadrupole. The dotted horizontal lines represent the expected isotropic re-scaling by the volume of the survey, $[D_V({\bf \Omega})/D_V({\bf \Omega}=\{0.31,0,69\})]^3$. The numerical values for the rescaling of these two (and other) cosmologies are displayed in Table~\ref{tab:scaling_cov_factors}.

\begin{table}[htbp]
    \centering
    \begin{tabular}{|c|c|c|c|}
    \hline
        $\Omega_m^{\rm fid}$ & $\Omega_\Lambda^{\rm fid}$ & $\Omega_k^{\rm fid}$ & $[D_V({\bf \Omega}^{\rm fid})/D_V({\bf \Omega}^{\rm ref})]^3$  \\
        \hline
        \hline
        0.16  & 0.69 & $+0.15$ & 1.170 \\
        0.21 & 0.69 & $+0.10$ & 1.108 \\ 
        0.26 & 0.69 & $+0.05$ & 1.051 \\ 
        0.31 & 0.69 & $0.00$ & 1 \\ 
        0.36 & 0.69 & $-0.05$ & 0.953 \\ 
        0.41 & 0.69 & $-0.10$ & 0.910 \\ 
        0.46  & 0.69 & $-0.15$ & 0.870 \\
        \hline
        0.31 & 0.54 & $+0.15$ & 0.892 \\ 
        0.31 & 0.59 & $+0.10$ & 0.925 \\ 
        0.31 & 0.64 & $+0.05$ & 0.961 \\    
        0.31 & 0.69 & $0.00$ & 1 \\ 
        0.31 & 0.74 & $-0.05$ & 1.042 \\ 
        0.31 & 0.79 & $-0.10$ & 1.087 \\
        0.31 & 0.84 & $-0.15$ & 1.136 \\ 
        \hline
    \end{tabular}
    \caption{Re-scaling isotropic factors at the eBOSS LRG sample effective redshift, $z=0.698$, for different choices of the fiducial cosmology parameters, $\Omega_m^{\rm fid},\,\Omega_\Lambda^{\rm fid},\,\Omega_k^{\rm fid}$, relative to the nominal (or reference) case $\{\Omega^{\rm ref}_m,\,\Omega^{\rm ref}_\Lambda\}=\{0.31,\,0.69\}$.}
    \label{tab:scaling_cov_factors}
\end{table}

From Fig.~\ref{fig:rescaling_comparison} we can conclude that the observed deviation is very well described by the isotropic volume rescaling factor of Table~\ref{tab:scaling_cov_factors}.

\begin{figure}[htb]
    \centering
    \includegraphics[scale=0.2]{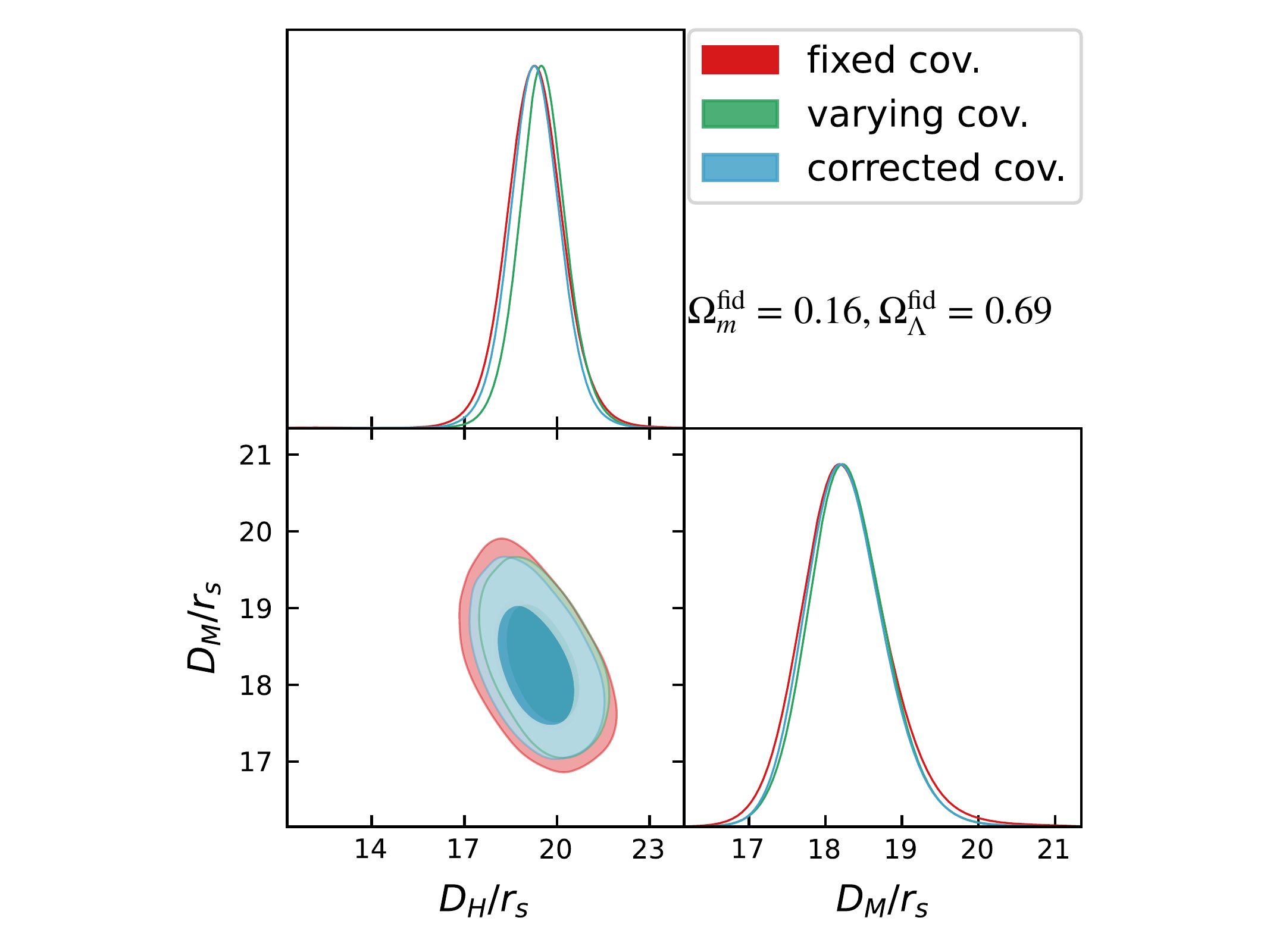}
        \includegraphics[scale=0.2]{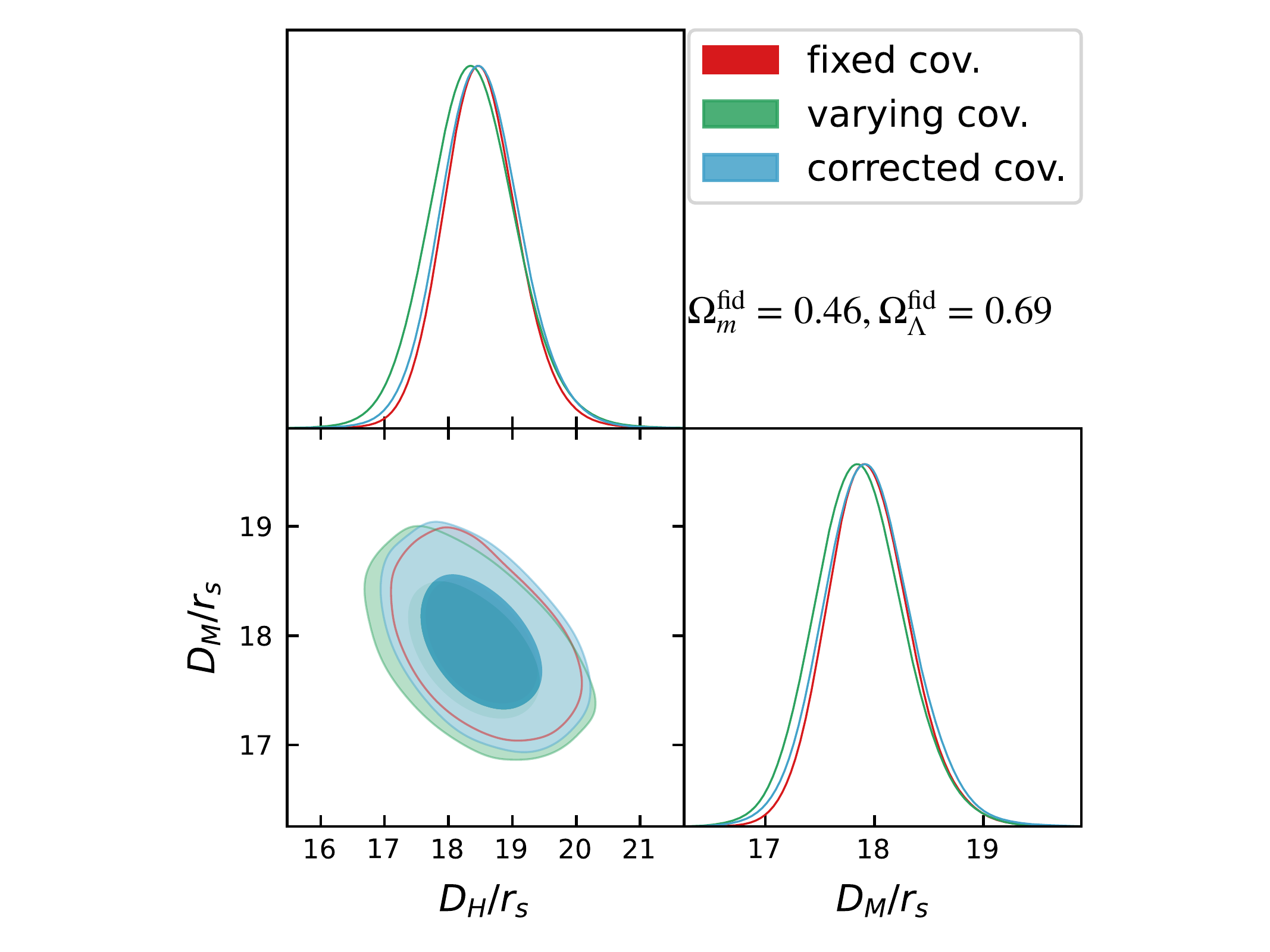}

\includegraphics[scale=0.2]{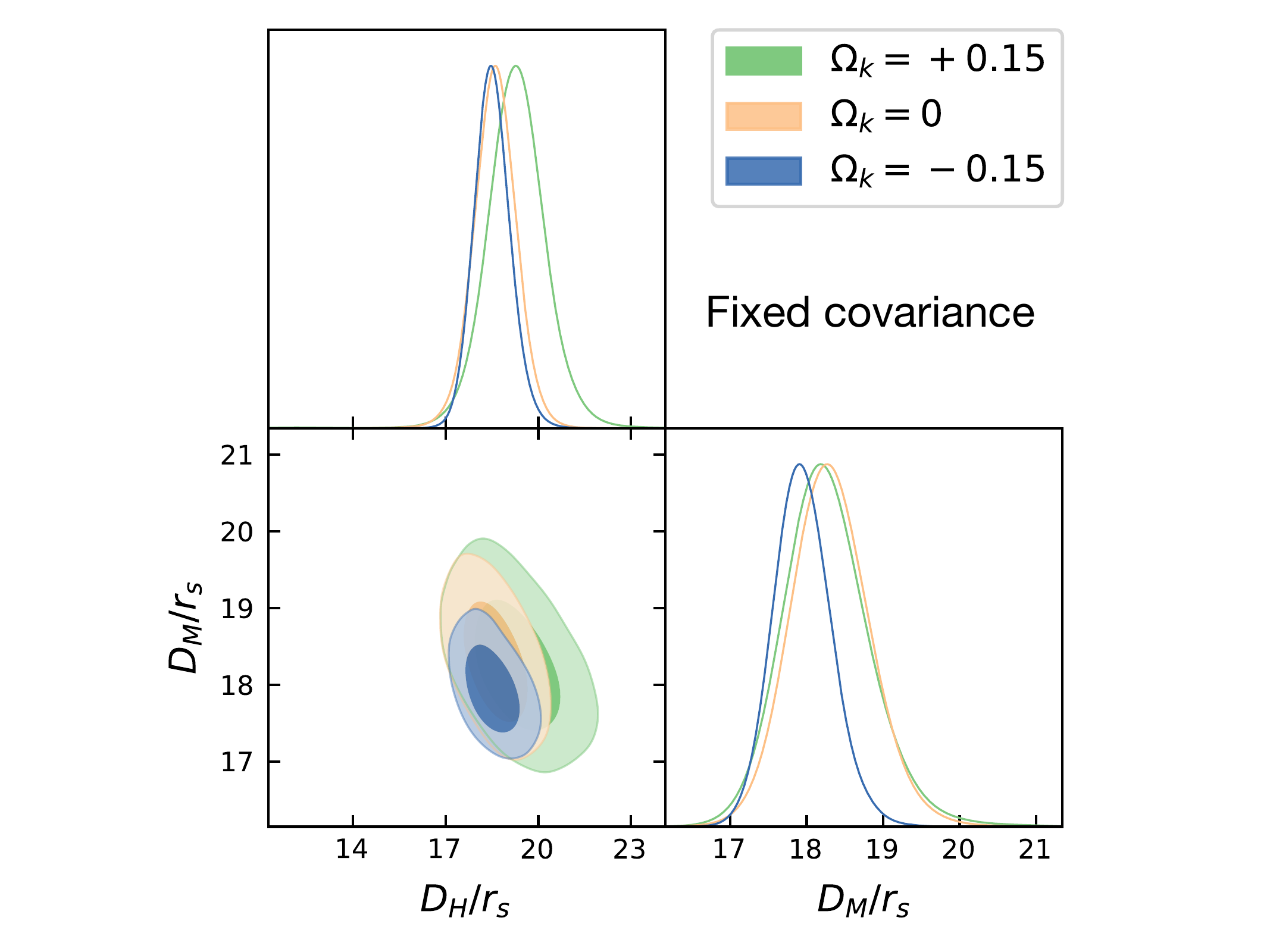}
\includegraphics[scale=0.2]{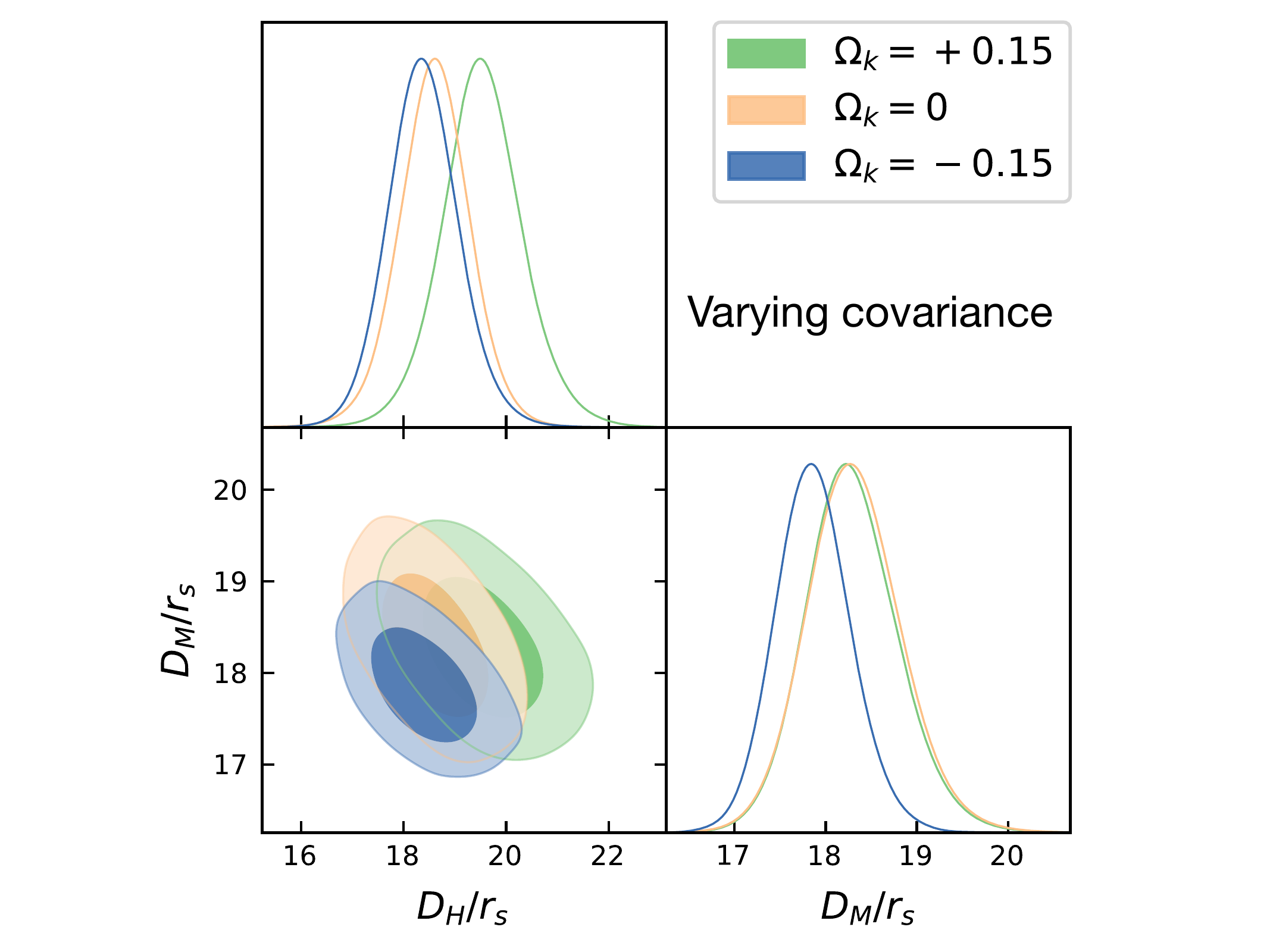}
        
    \caption{Posteriors for eBOSS LRG data in the range $0.6<z<1.0$ analyzed under different fiducial catalog cosmology values for transforming redshift into distances and template, using different prescriptions for inferring the covariance from mocks. In the top panels, the red contours represent the posteriors obtained by using the covariance computed from the power spectra of the mocks at the baseline case of $\{\Omega^{c}_m,\,\Omega_\Lambda^{c}\}=\{0.31,\,0.69\}$, the green contours for the covariance inferred from mocks computed at the same cosmology as the data (correct case, but computationally more expensive as requires re-computing 1000 \textsc{ezmocks} realizations per galactic cap for each fiducial case), and the case for the covariance inferred from the power spectra computed at the baseline cosmology, but re-scaling by the isotropic factors of Table~\ref{tab:scaling_cov_factors}. The bottom panels display the same data, but with the contours arranged by the type of covariance being used, which displays the over- and under-estimation when the covariance is not recomputed at each fiducial cosmology case.}
    \label{fig:contours_comp}
\end{figure}

The effect of the covariance rescaling on the inferred BAO distances is illustrated in  Fig.~\ref{fig:contours_comp}.

The top panels show two examples of adopted fiducial cosmologies (both for template and catalog) as indicated in the legend. The green contours are obtained by adopting the corresponding fiducial catalog cosmology for the data and the mocks used to compute the covariance (i.e., correct, exact treatment). The red contours are obtained by using a covariance matrix calculated from mocks where the catalog cosmology is the baseline one (\{$\Omega_m,\,\Omega_\Lambda\}=\{0.31,0.69\}$ i.e., different from the fiducial cosmology adopted for the data). The blue contours are obtained using the covariance from the baseline case with a rescaling of all the matrix elements by the factors of Table~\ref{tab:scaling_cov_factors}.
The bottom panels display the same information but compare cases with fixed covariance on the left  (corresponding to the red contours of the top panels), and cases with varying covariance in the bottom right panel (corresponding to the green contours in the top panels).

We see that ignoring the effect of changing the fiducial (catalog) cosmology of the mocks for covariance significantly under- and over-estimates the errors for the range of fiducial cosmologies explored in this paper, $-0.15<\Omega_k<+0.15$. Fully accounting for this change keeps the errors unaffected by the change of fiducial cosmology, supporting the assumption that the derived errors on $D_H/r_s$ and $D_M/r_s$ are independent of the assumption of fiducial cosmology. Also, this effect can be efficiently mitigated at a lower computational cost (without having to re-compute all the mocks at a new covariance choice) by correcting the covariance by the isotropic rescaling factor of Table~\ref{tab:scaling_cov_factors}.

\section{The difference between $r_d^I$ and $r_d^T$ and effect on BAO interpretation}
\label{sec:appC}

Motivated by the results of Ref.~\cite{Lewisrd2014}  we start by assuming that  $r_d^{T}\ne r_d^{I} $  and find their relation with the measured and theoretically predicted $\alpha_{\parallel,\perp}$ values. 
 On one hand, $r_d^I$ represents the drag epoch sound horizon scale computed from integrating the expansion history down to the drag epoch, 
\begin{equation}
\label{eq:rdI}
    r_d^I = \int^\infty_{z_d}\frac{c_s\,dz'}{H(z)},
\end{equation}
where $c_s$ is the speed of pressure waves in the photon-baryon plasma and $z_d$ stands for the epoch when the baryon optical depth is one (drag epoch). On the other hand, $r_d^T$ stands for the drag epoch sound horizon scale measured from the oscillatory template\footnote{For simplicity here we refer to the oscillatory template in $k$-space. Alternatively, one could measure the sound horizon scale from the oscillatory template in configuration space. We leave this approach for future work.}.

From the definition of $r_d^T$ the $\alpha_{\parallel,\perp}$ values obtained from the fit of the template to the data are a function of the {\it template} sound horizon scale as, 
\begin{equation}
\label{eq:ameasured}
 \alpha^{\rm meas.}=\frac{[D/r_d^{T}]^{\rm true}}{[D/r_d^T]^{\rm ref}}
\end{equation}
On the other hand, the $\alpha_{\parallel,\perp}$ that a cosmology inference code\footnote{At least the way it has been implemented so far in codes such as MontePython \cite{Montepython3} or Cobaya \cite{Cobaya}.} interprets as a function of $D_{H,M}/r_d$ are a function of the sound horizon scale obtained from the {\it integral} of Eq.~\ref{eq:rdI}, 
\begin{equation}
\label{eq:atheo}
    \alpha^{\rm th.}=\frac{[D/r_d^{I}]^{\rm true}}{[D/r_d^I]^{\rm ref}}.
\end{equation}
Hence,
\begin{equation}
\label{eq:atheo2}
    \alpha^{\rm th.}=\alpha^{\rm meas.}\left[ \frac{r_d^T}{r_d^I} \right]^{\rm true}\,\left[ \frac{r_d^I}{r_d^T} \right]^{\rm ref}.
\end{equation}
For the case of a reference cosmology very close to the true cosmology, the measured and theoretical $\alpha$ are the same; hence inference and interpretation are unbiased to each other. However, if the reference cosmology is different enough from the true underlying cosmology,  using  $\alpha^{\rm meas.}$ in the inference step {\it as if it were} $\alpha^{\rm th.}$ can potentially produce an uncontrolled systematic bias in the inferred parameters. Taking into account these two sound horizon scale distances in the inference cosmology pipeline is of key importance for BAO sub-percent precision pipelines. 

Here, since both reference and true cosmology are known, we can work out this bias (see appendix \ref{sec:appD} for how to measure $r_d^T$ in practice). 
In realistic applications, where the true underlying cosmology is not known,  $r_d^T$ needs to be computed at every step of the MCMC  during the cosmological inference (see \cite{Schoneberginprep}).
We can explicitly write this {\it a posteriori} correction as follows.

We rewrite  Eq.~\ref{eq:atheo} as,
\begin{equation}
\label{eq:atheo3}
    \left[\frac{D}{r_d^I}\right]^{\rm ref}\cdot  \alpha^{\rm th.} = \left[\frac{D}{r_d^I}\right]^{\rm true}.
\end{equation}
and substitute $\alpha^{\rm th.}$ by expression of  Eq.~\ref{eq:atheo2} obtaining
\begin{equation}    
\label{eq:ameasured2}
\left[\frac{D}{r_d^I}\right]^{\rm ref}\cdot  \alpha^{\rm meas} \left[ \frac{r_d^T}{r_d^I} \right]^{\rm true}\cdot\left[ \frac{r_d^I}{r_d^T} \right]^{\rm ref} = \left[\frac{D}{r_d^I}\right]^{\rm true}\equiv  \left[\frac{D}{r_d'}\right]^{\rm ref}  \alpha^{\rm meas},
\end{equation}
where the last identity defines,
\begin{equation}
\label{eq:rdproxy}
[r_d']^{\rm ref}\equiv [r_d^T]^{\rm ref}\left[\frac{r_d^I}{r_d^T}\right]^{\rm true}.
\end{equation}

Note that the rhs equality of Eq.~\ref{eq:ameasured2} is  fundamentally different from  the usual assumption done in BAO analyses and inference and what we have been using in Fig.~\ref{fig:exp_his_10mocks}, \ref{fig:inference_av10} and the empty symbols of  Fig.~\ref{fig:reanalysis}:

\begin{eqnarray}
{\rm Main\,\,text\,\,\,\,\,} & & {\,\,\,\,\,\rm Eq.\, C.6} \\ \nonumber
         \left[\frac{D}{r_d^I}\right]^{\rm ref}\cdot  \alpha^{\rm meas} = \left[\frac{D}{r_d^I}\right]^{\rm true}&\longrightarrow & \left[\frac{D}{r_d'}\right]^{\rm ref}\cdot  \alpha^{\rm meas} = \left[\frac{D}{r_d^I}\right]^{\rm true}        
\end{eqnarray}

Eq.~\ref{eq:ameasured2} tells us how to correctly interpret $\alpha^{\rm meas.}$ in terms of the angular and transverse distances given a reference cosmology: in the inference step of the standard analysis  $[r'_d]^{\rm ref}$ should be used instead of $[r_d^I]^{\rm ref}$. 

\section{Residual cosmology dependence of the sound horizon scale: measuring $r_d^T$}
\label{sec:appD}
As described in \cite{Lewisrd2014}, \S~\ref{sec:iterative}, and appendix~\ref{sec:appC}, the measured BAO shift in the oscillatory template, $\pazocal{O}_{\rm lin}(k)$, namely $r_d^T$, has a weak (but measurable) shift with respect to the early-universe quantity $r_d^I$.  
This shift is cosmology-dependent, which, if ignored,  can cause a significant systematic shift in the inferred cosmological parameters.
The findings presented in this appendix are further developed in \cite{Schoneberginprep}.

\begin{figure}
    \centering
    \includegraphics[scale=0.4]{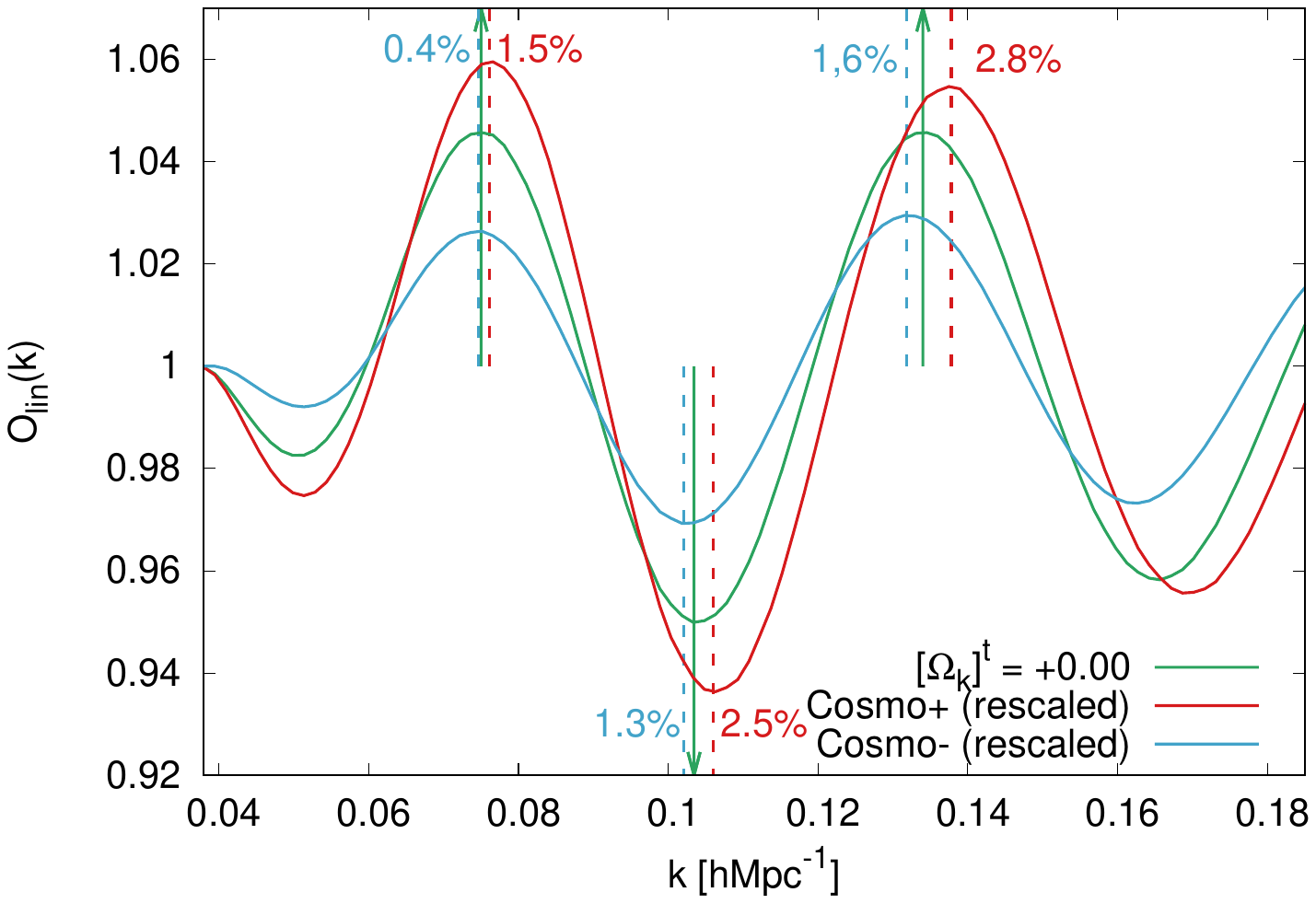}
    \caption{Oscillatory feature in the power spectrum $\pazocal{O}_{\rm lin}$ function (see Eq.~\ref{eq:Pbao} for the different cosmological models: green for the standard flat-$\Lambda$CDM, and red and blue for two extra cosmologies, Cosmo$\pm$, with curved universe displayed in Table~\ref{tab:bestfit}. The argument of the Cosmo$\pm$ cosmologies has been rescaled by the factor $[r^I_d]^{\Omega_k=0}/[r^I_d]^{{\rm Cosmo}\pm}$ to account for the different integral sound horizon scale values, $r_d^I$ (see text and Eq.~\ref{eq:rdI}). The vertical lines mark the maximum/minimum of the peaks/troughs for the 3 different $\pazocal{O}_{\rm lin}$ features, and the colored numbers present the per cent deviation with respect to the flat-$\Lambda$CDM for each of the peaks/troughs.}
    \label{fig:olins_rs}
\end{figure}

To visualize this effect we consider the 3 reference cosmologies  Cosmo$\pm$ (as in Table~\ref{tab:bestfit}) and the fiducial flat-$\Lambda$CDM cosmology (described at the beginning of \S~\ref{sec:methodology}). Taking as a reference the flat-$\Lambda$CDM cosmology, we re-scale the $\pazocal{O}_{\rm lin}(k)$ features of the  Cosmo$\pm$ cosmologies by the following factor in its argument,
\begin{equation}
    k\rightarrow  k \times \frac{[r_d^I]^{\Omega_k=+0.00}}{[r_d^I]^{{\rm Cosmo}\pm}}\equiv k^{\pm}
\end{equation}
and compare them with the $\pazocal{O}_{\rm lin}(k)$ of the $\Omega_k=0$ cosmology. If it were that  $r_d^T=r_d^I$, we would expect a perfect alignment of the peaks of the 3 $\pazocal{O}_{\rm lin}(k)$, despite having different amplitudes or shapes for the BAO oscillatory features. 
This  is illustrated  in Fig.~\ref{fig:olins_rs}, where the green curve displays $\pazocal{O}_{\rm lin}^{\Omega_k=0}(k)$, the red curves  
$\pazocal{O}_{\rm lin}^{{\rm Cosmo}+}(k^+)$
and the blue curves $\pazocal{O}_{\rm lin}^{{\rm Cosmo}-}(k^-)$. The dashed vertical lines mark the peaks and troughs of the oscillatory features corresponding to the cosmologies Cosmo$\pm$. The solid green arrows represent the same but for the reference flat-$\Lambda$CDM cosmology. The colored numbers report the percentage deviation with respect to the flat-$\Lambda$CDM cosmology, for each of the peaks and troughs. 

We report a shift which is significantly higher for the Cosmo$+$ than for the Cosmo$-$ cosmology, and that in both cases increases as a function of the peak/troughs. This effect is consistent with a {\it phase shift} that has been exploited already to extract information on the effective number of relativistic species, $N_{\rm eff}$ \cite{baumann_shift}. 

The net effect on the cosmological inference of \S~\ref{sec:iterative} is a 1.3\% increase in the sound horizon scale of Cosmo$+$ and a $0.1\%$ decrease for Cosmo$-$. This yields, for the two cosmologies Cosmo${\pm}$ the $r'_d$ values reported in Table~\ref{tab:bestfit}.

In summary, here we have shown that for non-flat cosmologies, this phase shift can cause residual cosmology-dependent effects that must be taken into account to avoid systematic shifts in the inferred cosmological parameters.

\section{Effect of the catalog fiducial cosmology on the survey window function}
\label{app:mask}
In this section, we aim to test the effect of the catalog fiducial cosmology in the window selection function. The main results of this paper rely on the assumption that, for the BAO scaling parameters, $\alpha_\parallel$ and $\alpha_\perp$ do not depend on whether the window selection function is computed at the flat fiducial cosmology ($\Omega_k^{\rm c}=0$) or at any of the other curved fiducial cosmologies ($\Omega_k^{\rm c}=\pm 0.15$), for the volumes studied here, either for BOSS+eBOSS volume or 10 times this volume. Working under this assumption saves us a lot of computational time because we avoid having to re-run the random-pair counts to compute the $W^2_{\ell}(s)$ functions (see appendix D of \cite{gilmarin20}). 

To test this assumption, we relax this condition and infer the BAO distance parameter in a fully consistent way, using the same fiducial cosmology for both the catalog redshift-to-distance conversion, and to infer the  $W^{(\ell)}(s)$ functions (what we refer as the $\Omega_k^{\rm mask}$ cosmology). For simplicity we only focus on the QSO sample, we apply a covariance of 100 times the BOSS+eBOSS mock data (recall that the data vector has also been generated in this case as 100 times the volume covered by the BOSS+eBOSS samples), and we keep fixed the cosmology of the template to the fiducial flat in all cases, $\Omega_k^{\rm t}=0$, as we want to isolate the effects on the catalog (and mask) cosmology only.

\begin{figure}
    \centering
    \includegraphics[scale=1]{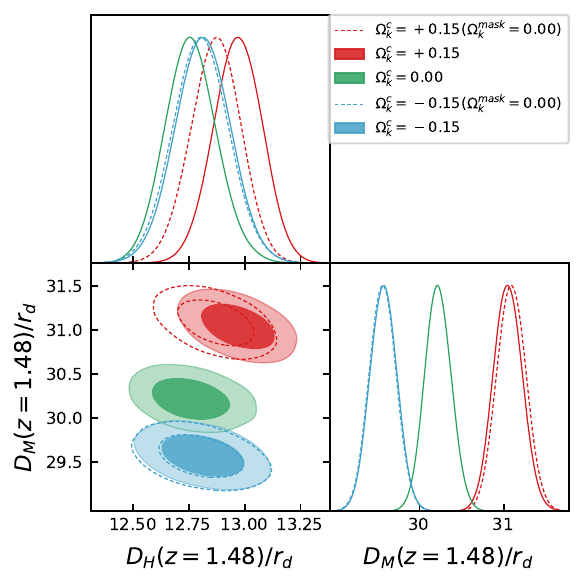}
    \caption{Impact of the selection window function cosmology on the QSO sample, $0.8<z<2.2$. The posteriors correspond to the BAO distances in $r_d$ units of the mean of 100 eBOSS QSO mocks analyzed with the covariance computed from 1000 mocks realizations and being rescaled to the 100 times eBOSS volume. Colors display the fiducial catalog cosmologies being used. The template cosmology has been fixed to the flat one for all cases. The dashed empty posteriors display the results when the fiducial flat cosmology is used to infer the window matrix functions, $W_\ell^2(s)$ (see appendix D of \cite{gilmarin20}), whereas the filled contours display the results obtained when using the same cosmology for the catalog and the window matrix functions calculation, $\Omega_k^{\rm c} = \Omega_k^{\rm mask}$.}
    \label{fig:mask_effect}
\end{figure}

Fig.~\ref{fig:mask_effect} display this effect: in red/blue are the posteriors when the catalog cosmology is $\Omega_k^{\rm c}=\pm0.15$ and the green when is $\Omega_k^{\rm c}=0$, as labelled. In all cases, $\Omega_k^{\rm t}=0$. The non-filled red/blue contours represent those constraints obtained with the $W^2_{\ell}(s)$ mask corresponding to the $\Omega_k^{\rm c}=0$ cosmology, whereas the filled ones correspond to the mask applied from the $W^2_{\ell}(s)$ functions computed from their own $\Omega_k^{\rm c}$ cosmology. The effect of the window function is negligible, even at this unrealistic large volume (errors correspond to a volume of $\sim90\,[{\rm Gpc}h^{-1}]^3$). We only observe $\sim0.5\sigma$ shift on $D_H/r_d$ for the $\Omega_k^{\rm c}=+0.15$. Therefore we conclude that this effect is negligible for volumes of $\sim 30\,[{\rm Gpc}h^{-1}]^3$, which validates the approximations performed in the main test of this paper. 

In addition, we also report the differences in $D_M/r_d$ for the 3 different cosmologies. This is in line with the result of Fig.~\ref{fig:exp_his_10mocks} (although in that figure also the template cosmology was varied). This supports the statements made in \S\ref{sec:geo} about the cause of the remaining systematics on $\Omega_m$ and $\Omega_k$ were not caused solely by the template cosmology, but have their origin in the catalog cosmology. As explained in \S~\ref{sec:geo} these differences can come from the wideness of the QSO redshift bin, where the approximation $|\Delta D_{H,M}| = |D_{H,M}(z_1)-D_{H,M}(z_2)|\ll D(z_{\rm eff})$ (where $z_{\rm eff}$ is the effective redshift of the bin, and $z_1$ and $z_2$ are the redshifts at the edges of the bin, $z_{\rm eff}=1.48$, $z_1=0.8$ and $z_2=2.2$), does not hold anymore for the precision displayed here.

\bibliographystyle{JHEP}
\bibliography{BAOcurvature}

\end{document}